\documentclass[11pt,a4paper,notoc]{article}
\pdfoutput=1
\usepackage{jcappub}
\usepackage{slashed}
\DeclareMathOperator\erf{erf}
\title{Quantifying (dis)agreement between direct detection experiments in a halo-independent way}

\arxivnumber{OUTP-14-14P}

\author{Brian Feldstein}
\author{and Felix Kahlhoefer}
\affiliation{Rudolf Peierls Centre for Theoretical Physics, University of Oxford, 1 Keble Road, Oxford OX1 3NP, United Kingdom}

\emailAdd{brian.feldstein@physics.ox.ac.uk}
\emailAdd{felix.kahlhoefer@physics.ox.ac.uk}
\date{\today}
\abstract{
We propose an improved method to study recent and near-future dark matter direct detection experiments with small numbers of observed events.  Our method determines in a quantitative and halo-independent way whether the experiments point towards a consistent dark matter signal and identifies the best-fit dark matter parameters.
To achieve true halo independence, we apply a recently developed method based on finding the velocity distribution that best describes a given set of data.
For a quantitative global analysis we construct a likelihood function suitable for small numbers of events, which allows us to determine the best-fit particle physics properties of dark matter considering all experiments simultaneously.
Based on this likelihood function we propose a new test statistic that quantifies how well the proposed model fits the data and how large the tension between different direct detection experiments is. We perform Monte Carlo simulations in order to determine the probability distribution function of this test statistic and to calculate the $p$-value for both the dark matter hypothesis and the background-only hypothesis.
}

\begin{document}

\maketitle

\section{Introduction}

The interpretation of results from dark matter (DM) direct detection experiments is significantly affected by uncertainties in the local DM density and velocity distribution~\cite{Kamionkowski:1997xg, Green:2002ht, Fairbairn:2008gz, Drees:2008bv, MarchRussell:2008dy, Kuhlen:2009vh, Green:2010gw, Strigari:2009zb, McCabe:2010zh, Green:2011bv, Fairbairn:2012zs,Strigari:2012gn}. While the assumed value for the local DM density only affects the determination of the overall DM scattering cross section, the 
assumed halo velocity distribution $ f(\mathbf{v})$ enters in the predicted event rate in a complicated way and changes the shape of the predicted recoil spectrum. Indeed, a nuclear recoil of energy $E_\text{R}$ can originate from any DM particle which has a speed greater than a minimum value $v_\text{min}(E_\text{R})$, and  experiments are then sensitive to the velocity integral $g(v_\text{min}) = \int_{v>v_\text{min}}   \hspace{-1mm} f(\mathbf{v}) / v \, \mathrm{d}^3\mathbf{v} $. 

In a recent publication~\cite{Feldstein:2014gza}, the authors of the present work proposed a new way to
treat halo uncertainties, which is based on determining the velocity integral that best describes a given set of data.  In that work it was shown that such a method can be applied to the task of determining DM parameters from the clear detection of a DM signal across multiple future experiments without 
requiring any assumptions about the underlying  velocity distribution. In the present work we demonstrate that a similar method can also be used to study less robust detections, with smaller statistics and with experiments that may appear to be in disagreement. Our goal here will be to determine, in a quantitative and halo independent way, to what extent such a set of data points towards a consistent DM signal.

Previous halo independent analyses for low statistics have appeared in the literature, but our approach improves upon them in two key respects:
\begin{itemize}
\item{{\bf True halo independence.} By virtue of using the method from~\cite{Feldstein:2014gza} to find the best-fit halo for a given set of data and DM parameters, our method yields results which are manifestly independent of any halo assumptions. The halo velocity integral is written as a sum of step functions with step heights optimized to fit the data. The number of steps is then taken to approach infinity, yielding the function $g(v_{\rm min})$  which matches the data as well as possible. This approach is similar to the one proposed in~\cite{Fox:2014kua} for the qualitative analysis of individual direct detection experiments, but should be contrasted with the quantitative global analyses in~\cite{Fairbairn:2008gz, Arina:2011si, Arina:2013jma} that make use of more restricted parameterizations of the halo. The halo-independent methods developed in~\cite{Peter:2009ak, Pato:2012fw, Peter:2011eu, Kavanagh:2012nr, Kavanagh:2013wba, Kavanagh:2013eya, Peter:2013aha}, on the other hand, do employ very general parameterizations of the velocity integral but only consider future data with high statistics. We will also discuss an extension to the method from~\cite{Feldstein:2014gza}, which allows for adding observational results on the Galactic escape velocity to the determination of the best-fit halo.}
\item{{\bf Quantitative global results for goodness of fit.} Since our method considers all experiments simultaneously when constructing the best-fit halo and determining DM model parameters, we are able to perform a global, quantitative analysis of how well the best-fit point describes the full set of data.
Previous halo-independent methods have typically only been able to compare experiments in pairs for fixed DM parameters~\cite{Fox:2010bz, Frandsen:2011gi, Gondolo:2012rs} (see also~\cite{HerreroGarcia:2011aa,HerreroGarcia:2012fu,Frandsen:2013cna,Bozorgnia:2013hsa,DelNobile:2013cva,DelNobile:2014eta, Scopel:2014kba}). The basic idea of these previous  approaches was to take experiments seeing a signal as giving measurements of the velocity integral $g(v_{\rm min})$, while experimental null results were taken to provide upper limits on this function.  If the measured values were seen to violate the upper limits anywhere in velocity space, the results were then deemed incompatible. 

Such analyses have several disadvantages.  In particular, as they only consider constraints from one experiment at a time, and at one point in velocity space at a time, they do not yield quantitative statements about the net amount of agreement or disagreement in the data. Even the direct comparison of two experiments is often only possible if the two experiments probe the same region of $v_\text{min}$-space~\cite{Fox:2010bz, McCabe:2011sr}.
Moreover, since for given DM parameters no aggregate test statistic is computed, it also follows that it is not possible for these methods to quantitatively compare different choices of DM parameters or to determine an allowed DM parameter region. Our approach, on the other hand, allows us to use all the information available in $v_\text{min}$-space in order to not only find the best-fit values for the DM parameters, but also to determine how well the proposed model actually fits the data.}
\end{itemize}

In addition to the halo optimization procedure of~\cite{Feldstein:2014gza}, our approach depends on three essential ingredients. The first is the construction of a likelihood function for each direct detection experiment, including those which report an excess as well as those which report null results. Depending on the available experimental information, we use either a binned likelihood function as in~\cite{Cowan:2010js} or an unbinned likelihood function~\cite{Barlow:1990vc}. The advantage over using statistical tests like the `maximum gap method'~\cite{Yellin:2002xd} is that we can multiply the individual likelihood functions in order to perform a global analysis and determine the best-fit model parameters for a given set of events. Similar constructions have been made in Ref.~\cite{Arina:2011si} in the context of Bayesian analyses of DM direct detection experiments and in Ref.~\cite{Kopp:2011yr} using a global $\chi^2$ function.

In contrast to a $\chi^2$ test statistic, the numerical value of the likelihood function at the best-fit point does not reveal how well the model is actually able to describe the data, because only \emph{likelihood ratios} between different points in parameter space are invariant under changes of variables in the experimentally observed quantities. Nevertheless, for ambiguous experimental hints and possible tension between different results it is of crucial importance to determine the goodness-of-fit of the best-fit point. We cannot simply perform a $\chi^2$ analysis in the present context, because we generally do not obtain sufficiently well populated bins of data from experiments observing a small number of events. The second additional ingredient for our approach is thus an alternate test statistic, based on likelihood ratios, which does indeed contain the desired goodness-of-fit information.

This statistic is given by dividing the global maximum likelihood by the maximum likelihood obtainable for each individual experiment on its own. Intuitively, this statistic measures how badly the experiments have to compromise in order to satisfy each other by moving to the global best-fit point.  Following~\cite{Maltoni:2003cu}, we will refer to this quantity as the ``parameter goodness of fit statistic" (PGS), with larger values of the statistic indicating larger disagreement. By considering both the PGS and the likelihood ratio between the best-fit point and the background-only hypothesis we can quantify the preference for a DM signal in a given set of experimental data. 

The final ingredient in our approach is to determine the probability distribution functions for our test statistics, in particular the probability that the PGS would have been observed to take a value as large as the measured one.  When the PGS is applied to data with high statistics, it can be shown analytically that
this quantity follows a $\chi^2$ distribution with an appropriate number of degrees of freedom~\cite{Maltoni:2003cu}. In the present case, however, we do not have such an analytic formula because of the small number of observed signal events, and thus the distribution of PGS values must be extracted from Monte Carlo (MC) simulations. It is thus fortunate that our halo optimization procedure from~\cite{Feldstein:2014gza} has the additional virtue of being numerically highly efficient, and is thus well suited to being implemented in MC simulations.

The structure of this paper is as follows. In Sec.~\ref{sec:basic} we briefly review the basic formalism of DM direct detection and introduce various ways to construct likelihood functions for direct detection experiments. In this context, we also review the halo-independent method that we employ. Sec.~\ref{sec:global} then deals with the combination of the individual likelihood functions into a global likelihood function, as well as possible test statistics that can be constructed from likelihood ratios. We also motivate the need for MC simulations to extract the distribution of these test statistics. We then present two specific examples in Secs.~\ref{sec:ex1} and~\ref{sec:ex2}, where we apply our method to recent results from direct detection experiments for different model assumptions. Finally, we discuss our findings in Sec.~\ref{sec:discussion}. Two appendices provide additional details on the experiments under consideration and the implementation of the MC simulations.

\section{The basic method}
\label{sec:basic}

In this section we provide the relevant formulas for the analysis of direct detection experiments and review the halo-independent method that we use. Moreover, we discuss how to construct a global likelihood function suitable for analyzing experimental hints of an emerging DM signal.

The minimum velocity that a DM particle must have in order to induce a nuclear recoil with energy $E_\text{R}$ is given by
\begin{equation}
v_\text{min}(E_\text{R}) = \sqrt{\frac{m_\text{N} E_\text{R}}{2 \mu^2}} \; ,
\end{equation}
where $\mu = m_\chi m_\text{N} / (m_\chi + m_\text{N})$ is the reduced mass of the DM-nucleus system and $m_\chi$ and $m_\text{N}$ are the DM mass and the mass of the target nucleus respectively. We can then define the velocity integral
\begin{equation}
g(v_\text{min}) = \int_{v > v_\text{min}} \frac{f(\mathbf{v})}{v} \,\text{d}^3\mathbf{v} \; ,
\end{equation}
where $f(\mathbf{v})$ is the local DM velocity distribution in the lab frame. Following~\cite{Fox:2010bz, Frandsen:2011gi}, we also introduce the rescaled velocity integral
\begin{equation}
\tilde{g}(v_\text{min}) = \frac{\rho \, \sigma_p}{m_\chi} g(v_\text{min}) \; ,
\label{eq:gtilde}
\end{equation}
where $\sigma_p$ is the DM-proton scattering cross section and $\rho$ is the local DM density.

Assuming elastic spin-independent scattering, we can then write the differential event rate with respect to recoil energy in a very simple form:
\begin{equation}
\frac{\text{d}R}{\text{d}E_\text{R}} =  
\frac{C_\text{T} (A,Z) \, F^2(E_{\text{R}})}{2 \, \mu_{n\chi}^2} \, \tilde{g}(v_\text{min}) \; ,
\label{eq:dRdE}
\end{equation}
where $F(E_\text{R})$ is the appropriate nuclear form factor and $\mu_{n\chi}$ is the reduced DM-nucleon mass. For the present study, we adopt the Helm form factor from~\cite{Lewin:1995rx}, in particular we neglect the small difference between proton and neutron form factor discussed in~\cite{Zheng:2014nga}. The coefficient $C_\text{T}(A, Z) \equiv \left[Z + f_n / f_p \, (A-Z)\right]^2$, where $A$ and $Z$ are the mass and charge number of the target nucleus, parameterizes the effective coupling to the entire target nucleus in terms of $f_n / f_p$, the ratio of DM-proton to DM-neutron couplings. If for a given isotope the coupling ratio satisfies $f_n / f_p \sim - Z / (A-Z)$, the effective coupling becomes very small due to destructive interference and scattering off this isotope is strongly suppressed~\cite{Kurylov:2003ra,Giuliani:2005my,Chang:2010yk,Feng:2011vu}.

In order to analyze direct detection experiments without making any ansatz for  the DM velocity distribution $f(\mathbf{v})$, we follow the approach outlined in~\cite{Feldstein:2014gza}, i.e.\ we parameterize the rescaled velocity integral $\tilde{g}(v_\text{min})$ in a fully general way and then find the form that best describes a given set of data. For this purpose, we divide the region of $v_\text{min}$-space under consideration into $N_\mathrm{s}$ intervals of the form $\left[v_j,\,v_{j+1}\right]$ and take the rescaled velocity integral to be constant ($\tilde{g}(v_\text{min}) = g_j$) within each interval. The fact that the velocity integral is both positive and monotonically decreasing leads to the requirement $0 \leq g_j \leq g_{j-1}$ for all $j$.

While formally an experiment with finite energy resolution is sensitive to all values of $v_\text{min}$, we only consider upward and downward fluctuations of at most $4\sigma$ when determining the relevant region of $v_\text{min}$-space. More precisely, we take the lowest observable recoil energy to be $E_\text{min} = E_\text{th} - 4 \Delta E(E_\mathrm{th})$, where $E_\text{th}$ is the lower threshold of the search window, and an analogous expression for $E_\text{max}$.\footnote{When studying the LUX experiment~\cite{Akerib:2013tjd} below, we follow the experimental collaboration and fix $E_\text{min} = 3\:\text{keV}$, i.e.\ we only include upward fluctuations from the energy range where the relative scintillation yield of liquid xenon has been measured.} We then take $v_0$ ($v_{N_\mathrm{s}}$) to be the minimum (maximum) value of $v_\text{min}(E_\text{min})$ ($v_\text{min}(E_\text{max})$) for all experiments. Note that in the examples we have considered, $v_0$ is always larger than zero, so that we do not constrain the velocity integral at very low velocities.

In order to calculate predicted event rates for a given direct detection experiment we need to know the probability $p(E_\mathrm{D}; E_\mathrm{R})$ that a DM interaction with true (i.e.\ physical) recoil energy $E_\mathrm{R}$ leads to a signal with detected energy $E_\mathrm{D}$. In terms of this function, and using our ansatz for $\tilde{g}(v_\text{min})$, we find
\begin{align}
\frac{\mathrm{d}R}{\mathrm{d}E_\mathrm{D}} & = \int_0^\infty \frac{\text{d}R}{\text{d}E_\text{R}} \, p(E_\mathrm{D}; E_\mathrm{R}) \, \mathrm{d}E_\mathrm{R} \nonumber \\
& = \sum_j g_j \, \frac{C_\text{T} (A,Z)}{2 \, \mu_{n\chi}^2} \int_{E_j}^{E_{j+1}} F^2(E_{\text{R}}) \, p(E_\mathrm{D}; E_\mathrm{R}) \, \mathrm{d}E_\mathrm{R} \nonumber \\
& \equiv \sum_j g_j \, D_j(E_\mathrm{D}) \; ,
\label{eq:Dj}
\end{align}
where $E_j$ is defined implicitly by $v_j = v_\text{min}(E_j)$ and we have introduced $N_\mathrm{s}$ functions $D_j(E_\mathrm{D})$ in the last step. If we are interested in the number of events $R_i$ predicted in a bin of the form $\left[E_i, E_{i+1}\right]$, we simply need to multiply $\mathrm{d}R/\mathrm{d}E_\mathrm{D}$ with the total exposure $\kappa$ and integrate over $E_\mathrm{D}$:
\begin{equation}
R_i = \kappa \int_{E_i}^{E_{i+1}} \frac{\mathrm{d}R}{\mathrm{d}E_\mathrm{D}} \, \mathrm{d}E_\mathrm{D} = \sum_j g_j \, \kappa \, \int_{E_i}^{E_{i+1}} D_j(E_\mathrm{D}) \, \mathrm{d}E_\mathrm{D} \equiv \sum_j D_{ij} g_j \; .
\end{equation}

In general, the function $p(E_\mathrm{D}; E_\mathrm{R})$ depends on the detection efficiency $\epsilon$ and the detector resolution. In the case where the detection efficiency depends only on $E_\mathrm{R}$ and the fluctuations can be approximated by a Gaussian distribution with standard deviation given by $\Delta E(E_\mathrm{R})$, we can write
\begin{equation}
p(E_\mathrm{D}; E_\mathrm{R}) = \epsilon(E_\mathrm{R}) \, \frac{1}{\sqrt{2 \pi}  \Delta E(E_\mathrm{R})}\exp\left[-\frac{(E_\mathrm{D} - E_\mathrm{R})^2}{2 \Delta E(E_\mathrm{R})^2} \right] \; .
\end{equation}
In this case, we can directly perform the integration over $E_\mathrm{D}$ and obtain
\begin{equation}
D_{ij} = \frac{\kappa \, C_\text{T} (A,Z)}{4 \, \mu_{n\chi}^2} \int_{E_j}^{E_{j+1}} F^2(E_{\text{R}}) \, \epsilon(E_\mathrm{R}) \, \left[\erf\left(\frac{E_{i+1} - E_\text{R}}{\sqrt{2} \Delta E_\text{R}}\right)-\erf\left(\frac{E_i - E_\text{R}}{\sqrt{2} \Delta E_\text{R}}\right)\right] \, \mathrm{d}E_\mathrm{R} \; ,
\end{equation}
where $\erf(x) = (2/\sqrt{\pi}) \int_0^x \exp(-t^2) \, \mathrm{d}t$.

The functions $D_j(E_\mathrm{D})$ and the constants $D_{ij}$ can easily be calculated numerically and depend only on the experimental details and the assumed DM properties, but are independent of astrophysics. The signal predictions then depend linearly on the constants $g_j$, making it easy to handle a large number of free parameters. In the following, we will always take $N_\mathrm{s} = 50$, 
and we have checked that a larger number of steps leads to only negligible improvements.

We now want to determine the constants $g_j$ that best describe given data from direct detection experiments. In contrast to the analysis performed in~\cite{Feldstein:2014gza}, we want to focus on current and near-future direct detection experiments. For this purpose, the $\chi^2$ test statistic introduced in~\cite{Feldstein:2014gza} is not appropriate, since it would place undue emphasis on bins where the number of predicted events is small, resulting in significant dependence on the choice of binning.  Moreover, in many cases it is desirable to use a method that requires no binning at all, as was recently suggested by~\cite{Fox:2014kua}. 

To study experiments with very small numbers of observed events we introduce a likelihood function of the form
\begin{equation}
 \mathcal{L} = \prod_\alpha \mathcal{L}^{(\alpha)} \; ,
\end{equation}
where $\alpha$ represents the different experiments under consideration. Depending on the available experimental information, we will either use a binned or an unbinned likelihood function for the individual likelihoods $\mathcal{L}^{(\alpha)}$. The binned likelihood function for an experiment with $n_\alpha$ bins is given by
\begin{equation}
-2 \log \mathcal{L}^{(\alpha)} = 2 \sum_{i=1}^{n_\alpha} \left[ R_i^{(\alpha)} + B_i^{(\alpha)} - N_i^{(\alpha)} + N_i^{(\alpha)} \log \frac{N_i^{(\alpha)}}{R_i^{(\alpha)} + B_i^{(\alpha)}} \right] \; ,
\label{eq:binnedL}
\end{equation}
where $R_i^{(\alpha)}$, $B_i^{(\alpha)}$ and $N_i^{(\alpha)}$ are the expected signal, expected background and observed number of events in the $i$th bin, respectively. For $N_i = 0$, the last term is taken to be zero.

The unbinned likelihood for an experiment with $N_\alpha$ observed events at energies $E_i^{(\alpha)}$ is given by
\begin{equation}
-2 \log \mathcal{L}^{(\alpha)} = -2 \sum_{i = 1}^{N_\alpha} \log \left(\frac{dR^{(\alpha)}}{dE_D}+\frac{dB^{(\alpha)}}{dE_D}\right)_{E_D = E_i} + 2 \left(R^{(\alpha)} + B^{(\alpha)}\right) \; ,
\label{eq:unbinnedL}
\end{equation}
where $dR^{(\alpha)}/dE_D$ and $dB^{(\alpha)}/dE_D$ are the predicted signal and background distribution, respectively, and $R^{(\alpha)} + B^{(\alpha)}$ is the total number of predicted events. Note that this function is not invariant under a change of units for the differential event rate. The absolute value of the unbinned likelihood function therefore carries no physical significance, and only (dimensionless) likelihood ratios can be studied in a sensible way.

Once we have constructed the individual likelihoods, we determine the step heights $g_j$ that minimize $-2 \log\mathcal{L}$, while at the same time satisfying the monotonicity requirement $0 \leq g_j \leq g_{j-1}$. It was pointed out in~\cite{Feldstein:2014gza} that this minimization is very simple for a $\chi^2$ test statistic, because any local minimum is automatically a global minimum. This property remains true when we replace the $\chi^2$ function by $\log \mathcal{L}$, making the minimization very fast even for a large number of parameters. We can therefore quickly determine the best possible form for the velocity integral given a set of data and DM model parameters.

\section{Studying the global likelihood function}
\label{sec:global}

The global likelihood function constructed in the previous section depends on the DM parameters (such as $m_\chi$ and $f_n / f_p$) and the parameters $\mathbf{g}$ that characterize the velocity integral.\footnote{Note that we have absorbed the DM scattering cross section into the definition of $\tilde{g}(v_\text{min})$. The likelihood function hence depends on $\sigma_p$ only implicitly via the halo parameters $\mathbf{g}$.} By finding the best-fit velocity integral, we obtain the profile likelihood function
\begin{equation}
\hat{\mathcal{L}}(m_\chi, f_n / f_p) = \max_{\mathbf{g}} \mathcal{L}(m_\chi, f_n / f_p, \mathbf{g}) \; .
\end{equation}
In the practical examples below, we will introduce additional nuisance parameters relating to background uncertainties and the cut-off velocity $v_\text{max}$. In these cases, the profile likelihood function is obtained by maximizing the likelihood with respect to both the halo parameters $\mathbf{g}$ and additional nuisance parameters.\footnote{In the presence of additional nuisance parameters it is no longer necessarily true that the likelihood function has only a unique minimum. In practice, this problem is addressed by first calculating a profile likelihood function by maximizing with respect to the halo parameters and then determining the optimum values for the nuisance parameters.} In other words, $\hat{\mathcal{L}}$ always only depends on the assumed particle physics parameters of DM.

It is now straightforward to find the DM parameters that maximize the profile likelihood function. We refer to these parameters as $\hat{m}_\chi$ and $\hat{f}_n / \hat{f}_p$ and call the maximum value of the profile likelihood function $\mathcal{L}_\text{max}$. Nevertheless, as long as the presence of a DM signal in the existing data is not established, this result in itself has very little meaning, since one would always obtain a best-fit point even for highly incompatible data sets. Rather than inferring best-fit parameter regions, we therefore want to address the following two questions: 1) How strongly is the best-fit DM interpretation of the data favored over the background-only hypothesis? 2) How well does the global best-fit model describe the data from the individual experiments? 

To answer the first question, we can calculate the likelihood for the background-only hypothesis, $\mathcal{L}_\text{bkg}$, by setting $\tilde{g}(v_\text{min}) = 0$ (and maximizing the likelihood with respect to all other nuisance parameters). We then define the likelihood ratio for the signal and background hypotheses: 
\begin{equation}
q_\text{bkg} = -2 (\log \mathcal{L}_\text{bkg} - \log \mathcal{L}_\text{max}) \; .
\end{equation}
Note that, by construction $q_\text{bkg} \geq 0$.\footnote{The background only hypothesis with $\tilde{g}(v_\text{min}) = 0$ is contained in the set of models which are optimised over to find $\mathcal{L}_\text{max}$.  This is why $q_\text{bkg}$ is necessarily non-negative.}

In order to quantify the $p$-value of the background-only hypothesis, we now need to quantify how likely it is to obtain a value of $q_\text{bkg}$ as large as the one implied by the data simply from random fluctuations of the expected backgrounds. Naively, $q_\text{bkg}$ should follow a $\chi^2$-distribution with number of degrees of freedom equal to the number of parameters fitted in order to obtain $\mathcal{L}_\text{max}$. In the present case, however, there are two additional complications: First of all, the notion of free parameters is rather unclear, since most of the parameters $g_j$ used to describe the velocity integral are not actually fitted but simply set to one of their boundary values (i.e.\ either $g_j = g_{j-1}$ or $g_j = 0$)~\cite{Feldstein:2014gza, Fox:2014kua}. Moreover, even if we kept the velocity integral fixed, $q_\text{bkg}$ would not necessarily follow a $\chi^2$-distribution for experiments with a small number of expected events. This complication can clearly be seen in the MC simulations of $q_\text{bkg}$ performed by the CDMS-II collaboration~\cite{Billard:2013gfa}.

In order to determine the probability distribution function of $q_\text{bkg}$ we therefore need to run MC simulations of the background distributions, i.e.\ we simulate the possible results of repeating the experiments under consideration many times in the absence of a DM signal. For each individual realization, we need to repeat the procedure outlined above, i.e.\ we find first $\hat{\mathcal{L}}$ and then $\mathcal{L}_\text{max}$ by determining the velocity integral and the DM parameters that best describe each simulated data set. The $p$-value of the background-only hypothesis is then defined as
\begin{equation}
p(\text{background}) = p(q_\text{bkg} > \bar{q}_\text{bkg}) \; ,
\end{equation}
where $\bar{q}_\text{bkg}$ is the actually observed value of $q_\text{bkg}$ in the published data.

Even if there is a strong preference for an additional signal contribution on top of the known backgrounds, this observation is not sufficient to conclude that the proposed best-fit DM model actually gives a good description of the data. For example, it is possible that (for the DM parameters under consideration) an excess in one experiment is in significant tension with the exclusion limit from another experiment. In this case, the global best-fit point will be a compromise between the two results, unable to fully explain the observed excess in the first experiment while still predicting an uncomfortably large number of events in the second.

To study the amount tension between individual experiments and their overall agreement with the model predictions, we introduce another test statistic in analogy to the parameter goodness-of-fit method discussed in~\cite{Maltoni:2003cu} (see also \cite{Kopp:2011yr}).  For a given set of data we define $\mathcal{L}_\text{max}^{(\alpha)}$ as the maximum likelihood for the experiment $\alpha$, obtained by varying both the halo parameters and the particle physics properties of DM.\footnote{Note that, for a single experiment with a single target element, $\hat{\mathcal{L}}^{(\alpha)}$ is typically independent of $m_\chi$ and $f_n / f_p$, since a change in the DM parameters can always be compensated by a change in the velocity integral. It is therefore trivial to determine $\mathcal{L}_\text{max}^{(\alpha)}$.} We then define $\mathcal{L}_\text{PG} = \prod \mathcal{L}_\text{max}^{(\alpha)}$. Note that, in contrast to our definition of $\mathcal{L}_\text{max}$, we allow different parameters for each experiment, i.e.\ we optimize the DM parameters and the velocity integral separately for each experiment. We can now define our second test statistic
\begin{equation}
q_\text{PG} = -2 ( \log \mathcal{L}_\text{max} - \log \mathcal{L}_\text{PG}) \; ,
\end{equation}
which by construction satisfies $q_\text{PG} \geq 0$.

Intuitively, $q_\text{PG}$ quantifies how much the individual experiments have to compromise in order to find a consistent description of all available data. If the individual experiments are in good agreement, we would expect that $q_\text{PG}$ is rather small, since similar parameters can fit all experiments simultaneously. A very large value of $q_\text{PG}$, on the other hand, would point towards strong disagreement between the individual experiments, because the respective best-fit parameters strongly disagree. To extract more quantitative statements about the probability distribution of $q_\text{PG}$ we again need to rely on MC simulations, but this time we want to generate events under the assumption of the best-fit DM model (rather than under the background-only hypothesis as above). We define the $p$-value of the best-fit point as
\begin{equation}
p(\text{best}) = p(q_\text{PG} > \bar{q}_\text{PG}) \; ,
\end{equation}
where, as before, $\bar{q}_\text{PG}$ denotes the actually observed value of $q_\text{PG}$.

We can generalize our definition of $q_\text{bkg}$ and $q_\text{PG}$ to apply not only at the best-fit point, but for any set of DM parameters:
\begin{align}
q_\text{bkg}(m_\chi, f_n / f_p) & = -2 \left[\log \mathcal{L}_\text{bkg} - \log \hat{\mathcal{L}}(m_\chi, f_n / f_p) \right] \nonumber \\
q_\text{PG}(m_\chi, f_n / f_p) & = -2 \left[\log \hat{\mathcal{L}}(m_\chi, f_n / f_p) - \log \mathcal{L}_\text{PG}\right] \; .
\end{align}
Our previous definitions then correspond to $q_\text{x} = q_\text{x}(\hat{m}_\chi, \hat{f}_n / \hat{f}_p)$. 
Note that the set of DM parameters which maximizes $q_\text{bkg}$ also minimizes $q_\text{PG}$.

In the following two sections, we will  make our discussion more explicit
 for a specific set of experimental results and different DM models. We emphasize that we study the currently available experimental data simply to illustrate our method. The aim of these examples is not to advertise a particular DM interpretation of any recently observed excesses. In fact, we will show that~-- under certain assumptions~-- a DM interpretation is entirely inconsistent with the data for any DM velocity distribution. Our objective is therefore to demonstrate how our method can be applied to future more reliable data sets.

\section{Example 1: CDMS-Si without isospin dependence}
\label{sec:ex1}

To give an example for the method introduced above and to determine the test statistics $q_\text{bkg}$ and $q_\text{PG}$ as well as their distributions in a specific case, we now turn to the analysis of several recent results from direct detection experiments. Our aim is to compare the excess claimed by CDMS-Si~\cite{Agnese:2013rvf} with the recent bounds from LUX~\cite{Akerib:2013tjd} and SuperCDMS~\cite{Agnese:2014aze} for different values of the DM mass.  In the subsequent section, we will allow the neutron-proton coupling ratio to vary, while here we set $f_n/f_p = 1$.

To study CDMS-Si, we follow the approach from Ref.~\cite{Frandsen:2011gi, Billard:2013gfa, Fox:2014kua} and use the extended maximum likelihood method from Eq.~(\ref{eq:unbinnedL}). The energies of the three observed events are $E_\text{R} = 8.2,\ 9.5\text{ and }12.3\:\text{keV}$. For LUX and SuperCDMS we have insufficient information on the background distribution and therefore use the binned maximum likelihood method for these two experiments. Details on the assumed detector properties and background distributions are provided in Appendix~\ref{ap:experiments}. Before combining the likelihood functions for a global analysis, we validate our approach by confirming that we recover the standard results from the literature using the individual likelihood functions.

\begin{figure}[t]
\centering
\includegraphics[width=0.42\textwidth, clip, trim = 0 -12 0 0]{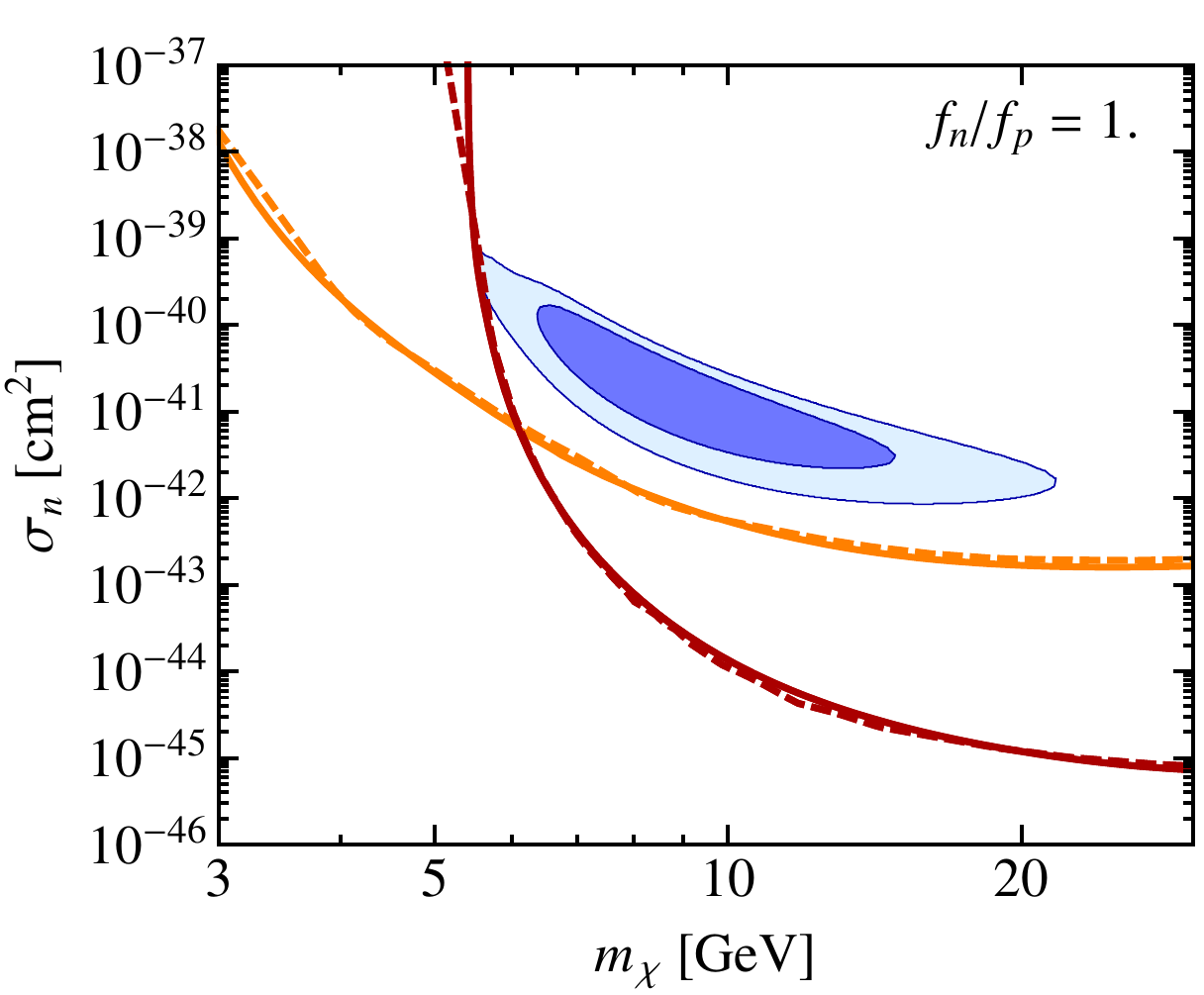}
\quad
\includegraphics[width=0.53\textwidth]{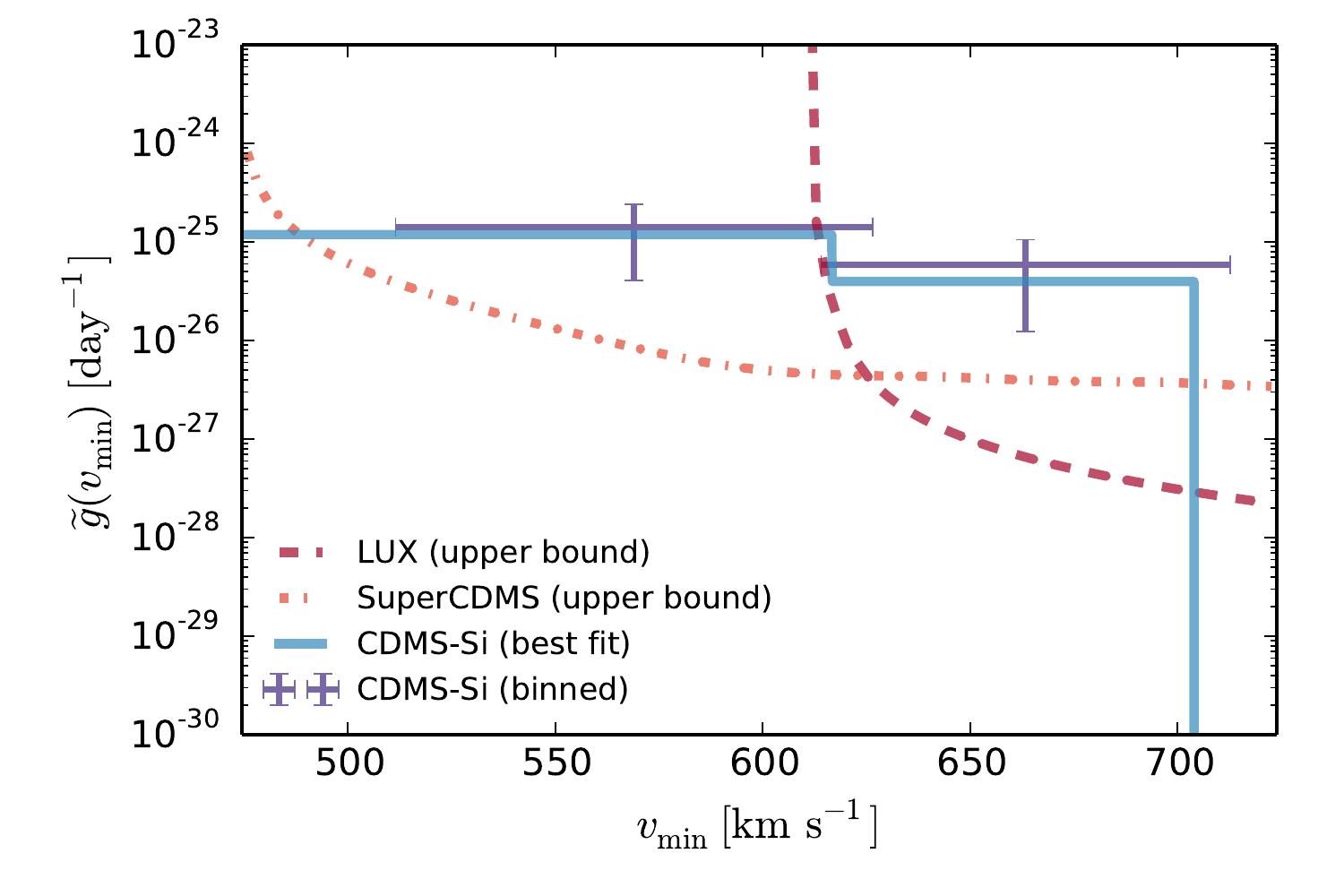}
\caption{Left: Best-fit DM mass and cross section for CDMS-Si and bounds from LUX and SuperCDMS assuming isoscalar DM and the Standard Halo Model velocity integral. Solid lines indicate the bounds obtained from our likelihood function, dashed lines indicate the published bounds. Right: Inferred values and bounds for the DM velocity integral assuming isoscalar DM with $m_\chi = 7$ GeV. Both plots agree well with the corresponding results in the literature.}
\label{fig:standard}
\end{figure}

The left panel of Fig.~\ref{fig:standard} shows the exclusion bounds (for LUX and SuperCDMS) and the CDMS-Si best-fit region in the $m_\chi$-$\sigma_p$ parameter plane under the assumption of the Standard Halo Model for the DM velocity distribution, which is a truncated Maxwell-Boltzmann distribution with velocity dispersion $\sigma_\text{dis}= \sqrt{3/2} \times 220$~km\,s$^{-1}$ and escape velocity $v_{\rm{esc}}=540$~km\,s$^{-1}$ in the Galactic rest frame. Our results agree well with the published bounds.

For the right panel of Fig.~\ref{fig:standard} we analyze the three experiments in $v_\text{min}$-space, following the method first introduced in~\cite{Fox:2010bz}. The basic idea of this method is that an experimental null result gives an upper bound on the rescaled velocity integral as a function of $v_\text{min}$. To obtain the bound at $v_\text{min} = \hat{v}_\text{min}$, one can simply use the standard techniques for setting an exclusion limit, but with
\begin{equation}
\tilde{g}(v_\text{min})= \tilde{g}(\hat{v}_{\rm{min}})\Theta(\hat{v}_{\rm{min}}-v_\text{min})\;,
\end{equation}
which is the velocity integral which gives the smallest DM signal consistent with monotonicity for a given  $\tilde{g}(\hat{v}_{\rm{min}})$.
The bounds we obtain for LUX and SuperCDMS agree well with the ones shown in~\cite{DelNobile:2014sja}.

An observed excess, on the other hand, can be used to deduce information on $\tilde{g}(v_\text{min})$ by binning both the observed events and the background expectation and then inverting Eq.~(\ref{eq:dRdE}) (see~\cite{Frandsen:2013cna}). For a bin width of $3\:\text{keV}$ one obtains the two data points shown in blue. To avoid the need for binning, it was suggested in~\cite{Fox:2014kua} to directly show the rescaled velocity integral that maximizes the unbinned likelihood function for CDMS-Si. Our best-fit velocity integral is indicated by the cyan line. This result agrees well with the one found in~\cite{Fox:2014kua}. 

It is important to note that we obtain the best-fit velocity integral for CDMS-Si in a way that is rather different from the one proposed in~\cite{Fox:2014kua}. It was shown~\cite{Fox:2014kua} that the unbinned likelihood is maximized by taking the velocity integral to be a sum of $n$ steps, where $n$ is smaller than or equal to the number of observed events. The best-fit velocity integral can therefore be found by simply maximizing the likelihood with respect to the height and position of each of these steps. In our approach we allow for a much larger number of steps, but find agreement with the results from~\cite{Fox:2014kua} in that the best-fit velocity integral only consists of a very small number of steps with non-zero step heights. Nevertheless, our approach allows us to determine the endpoints of these non-trivial steps in a numerically very efficient way (since varying step endpoints directly requires changing endpoints of integration in equation (\ref{eq:Dj}), and is therefore numerically taxing). This will become important when we combine the information from several experiments (as well as for the case of upcoming experiments with more than a few observed events).

Having validated the individual likelihood functions, we can now combine all the information to construct a global likelihood function and determine $\mathcal{L}_\text{max}$. We will first focus on the case of isoscalar DM with $f_p = f_n$ and then turn to different coupling ratios subsequently. For this assumption, the only free DM parameter is the DM mass $m_\chi$.

\begin{figure}[tb]
\begin{minipage}[t]{0.49\linewidth}
\centering
\includegraphics[width=0.99\textwidth]{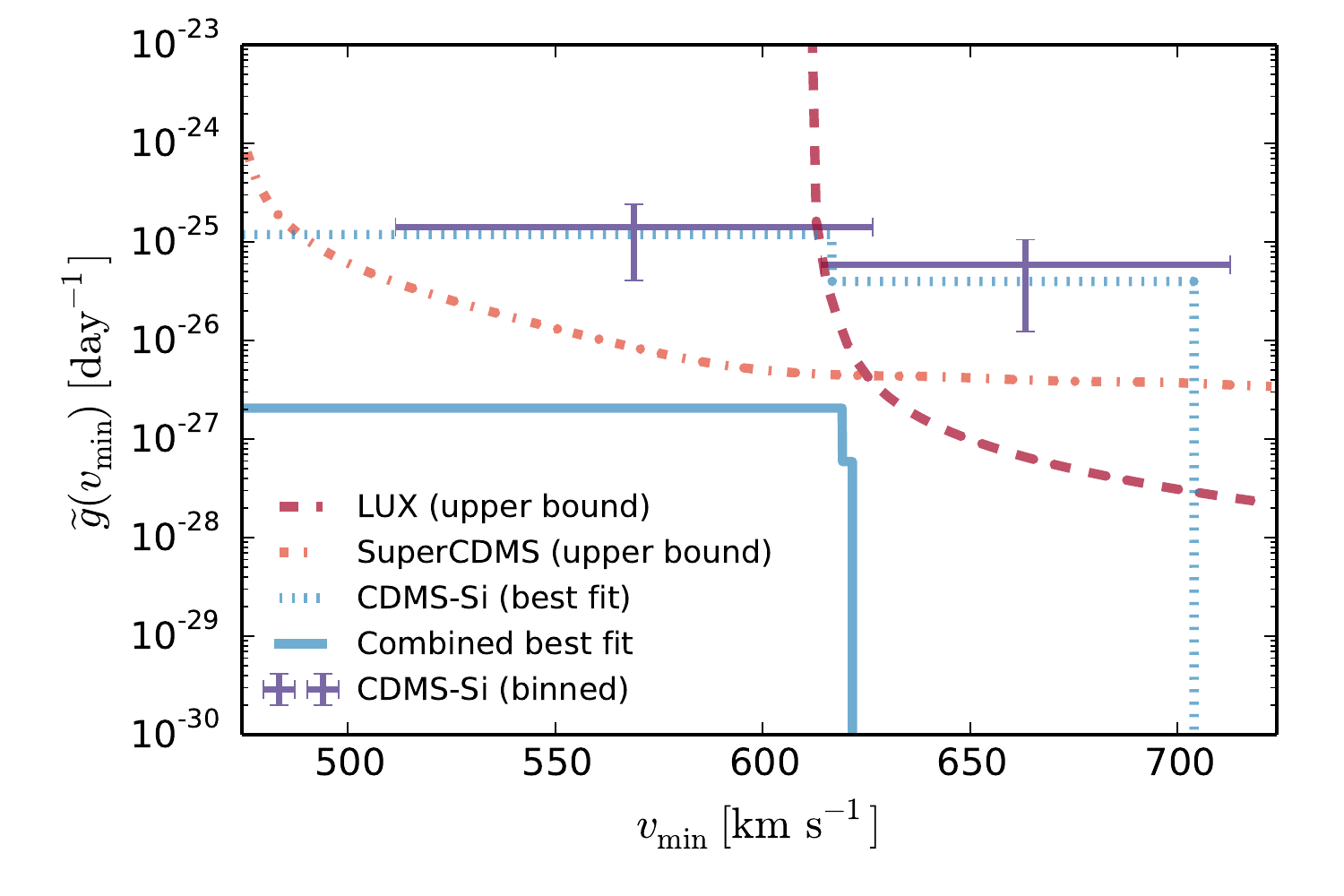}
\caption{Best-fit velocity integral obtained from the global likelihood function (solid line) compared to the individual bounds from LUX and SuperCDMS and the best-fit form inferred from CDMS-Si alone (dashed line) for $m_\chi = 7\:\text{GeV}$.}
\label{fig:combined}
\end{minipage}
\hspace{0.5cm}
\begin{minipage}[t]{0.47\linewidth}
\centering
\includegraphics[width=0.99\textwidth, clip, trim = 5 5 50 9]{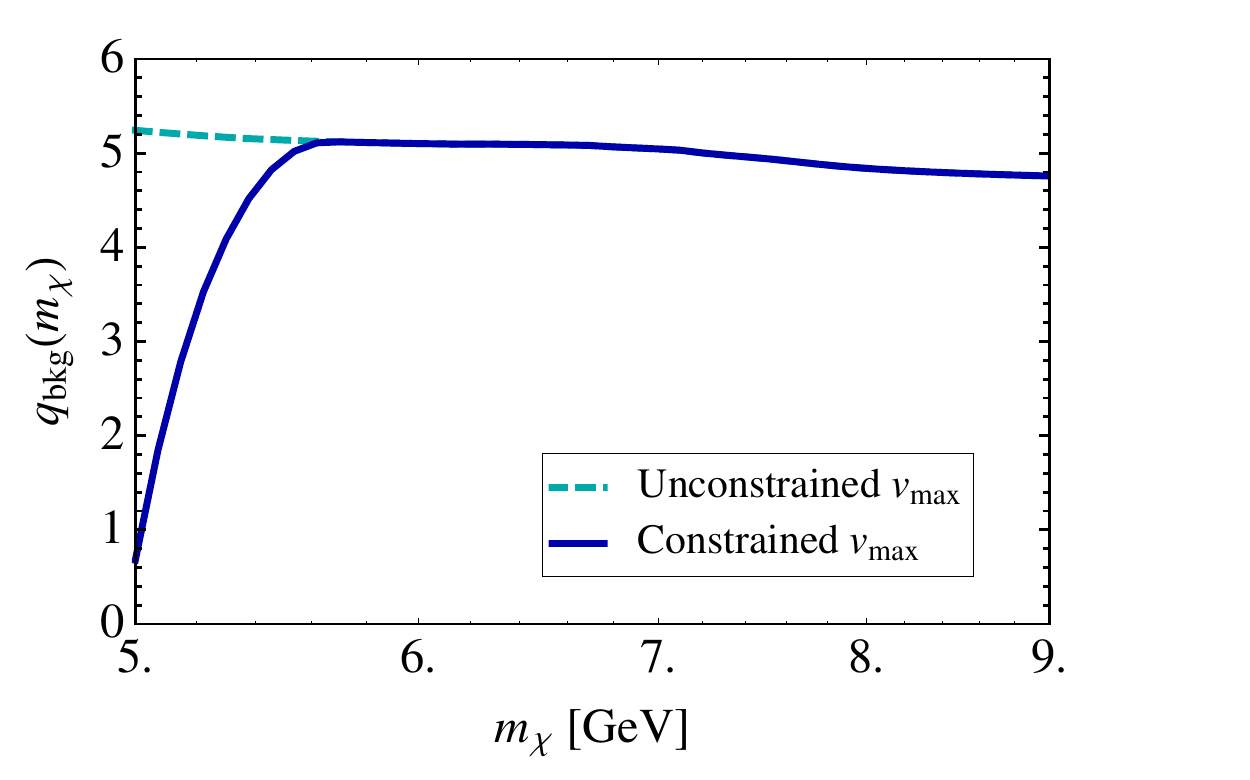}
\caption{The log likelihood ratio $q_\text{bkg}$ as a function of $m_\chi$ with (solid) and without (dashed) penalty for exceeding the escape velocity.}
\label{fig:escape}
\end{minipage}
\end{figure}

As expected from Fig.~\ref{fig:standard} for isoscalar DM, the constraints from LUX and SuperCDMS very strongly disfavor the best-fit interpretation of CDMS-Si. Fig.~\ref{fig:combined} shows the best-fit velocity integral obtained from the combined likelihood of all three experiments for $m_\chi = 7\:\text{GeV}$. As expected, this velocity integral respects the bounds on the velocity integral obtained from the individual likelihood functions. We observe that the combined best fit is about two orders of magnitude smaller than the best fit obtained for CDMS-Si alone. Although there is a significant preference of the best-fit point over the background-only hypothesis ($q_\text{bkg} = 5.0$), the obvious tension between the individual experiments is reflected in the fact that we obtain $q_\text{PG} = 12.7$. We will see below that these numbers correspond to $p$-values of about $1.5\%$ and $0.5\%$, respectively. In other words, while the background only hypothesis is strongly disfavored, the proposed model also does not give a good fit to the observed data

Let us now discuss whether a different choice of $m_\chi$ could give a better fit to the data and thereby increase $q_\text{bkg}$ and decrease $q_\text{PG}$. Indeed, since CDMS-Si has the lightest target element of the experiments under consideration, the disagreement between the three experiments can be reduced by making $m_\chi$ smaller and smaller.  This reduces the overlap region in $v_\text{min}$-space probed by the experiments, and allows for reducing tension by an appropriate choice of $\tilde{g}(v_\text{min})$.
Although doing so indeed leads to an increasing likelihood for the experimental data, there is a big price to pay: Moving to smaller masses pushes the best-fit velocity integral to be non-zero at higher and higher values of $v_\text{min}$. There are, however, strong observational constraints on the Galactic escape velocity and the benefit of going to lower masses is significantly outweighed by the resulting tension of the best-fit velocity integral with astrophysical bounds.

To check for a sensible DM interpretation, we clearly want to restrict ourselves to velocity distributions with reasonable escape velocities. To impose the observational constraints, we therefore introduce an additional term in our likelihood function:
\begin{equation}
 -2 \log \mathcal{L} \rightarrow -2 \log \mathcal{L} + \frac{(v_\text{max} - v_\text{esc} - v_\text{E})^2}{\Delta v^2} \; ,
\end{equation}
where $v_\text{max}$ is the maximum velocity $v_j$ for which $g_j \neq 0$ and we take $v_\text{esc} = 540$ km/s, $v_\text{E} = 225$ km/s and $\Delta v = \sqrt{\Delta v_\text{esc}^2 + \Delta v_\text{E}^2} \approx 42 \:\text{km/s}$ in accordance with~\cite{Smith:2006ym} and~\cite{Schoenrich:2009bx}. This extra term penalizes velocity integrals with exceedingly large velocities, treating the measured value of $v_\text{esc} + v_\text{E}$ as approximately following a normal distribution.\footnote{In practice we treat $v_\text{max}$ as an independent nuisance parameter, i.e.\ we impose $\tilde{g}(v_\text{min}) = 0$ for $v_\text{min} > v_\text{max}$ when determining the best-fit velocity integral numerically, and then repeat the optimization for different values of $v_\text{max}$. This procedure is necessary because there are typically several local minima in the full $\left\{g_j, v_\text{max}\right\}$ halo parameter space.  Recall that this does not occur when only the $g_j$ step parameters are used.}

\begin{figure}[t]
\centering
\includegraphics[width=0.46\textwidth]{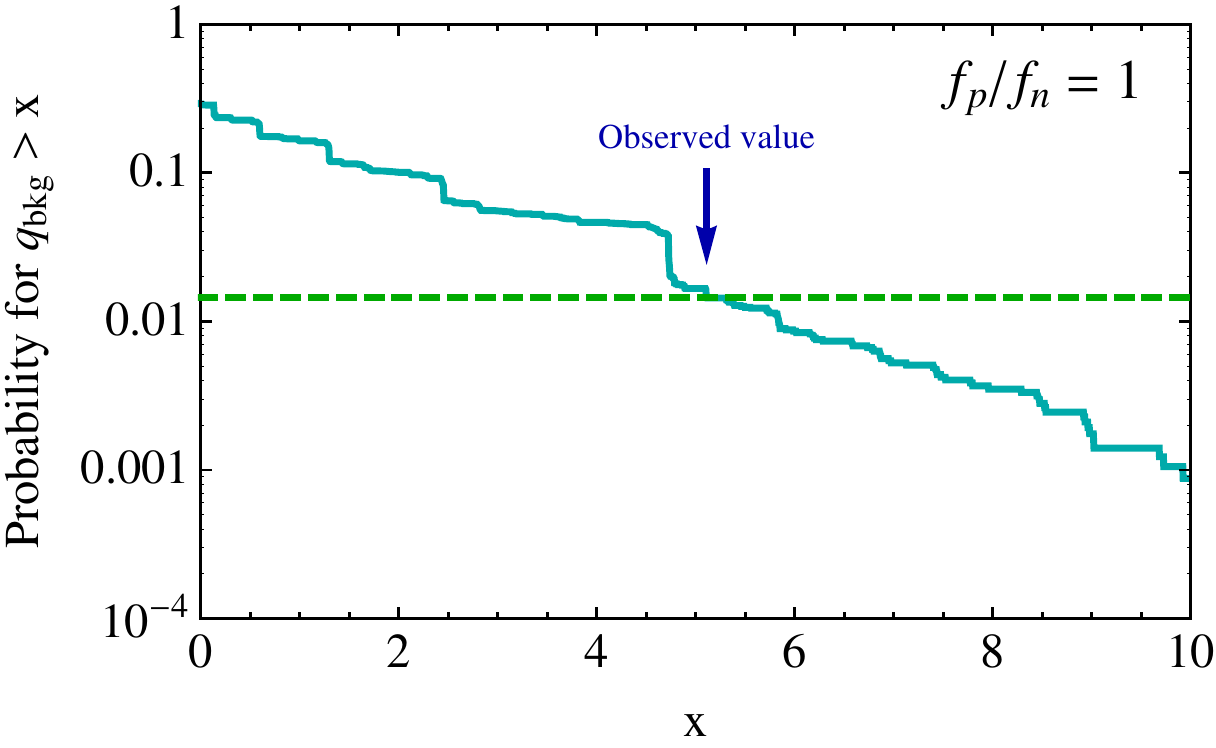}
\quad
\includegraphics[width=0.46\textwidth]{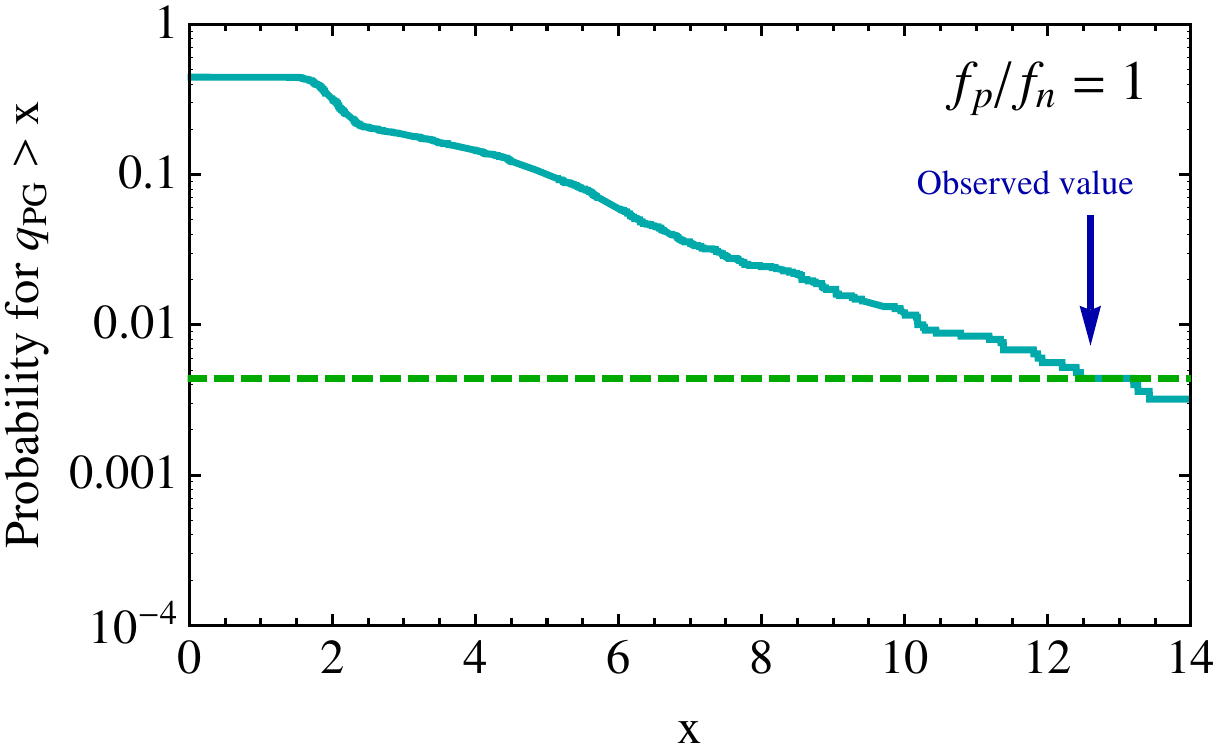}
\caption{$p$-values for $q_\text{bkg}$ based on the background-only hypothesis (left) and for $q_\text{PG}$ based on the best-fit DM model.}
\label{fig:MC1}
\end{figure}

Indeed, with this additional term it becomes unfavorable to reduce the DM mass significantly below $6\:\text{GeV}$ (see Fig.~\ref{fig:escape}), because the two events in CDMS-Si at higher energies can no longer be explained by DM with a reasonable escape velocity. The best-fit point is found for $m_\chi = 5.7\:\text{GeV}$ and gives $q_\text{bkg} = 5.1$ and $q_\text{PG} = 12.6$.\footnote{With no constraint on $v_\text{esc}$, and for e.g. $m_\chi = 3$~GeV, we find
 $q_\text{bkg} = 6.2$ and $q_\text{PG}  = 11.5$, though very large velocities with  $v_\text{max} \sim 1430$~km/s are required.} Clearly, significant tension remains even for the optimum choice of $m_\chi$.
 To determine how large this tension actually is, we now run MC simulations of the background distribution (to determine the $p$-value of the background-only hypothesis) as well as MC simulations of the best-fit DM model (to determine the $p$-value of the DM interpretation). Our results are shown in Fig.~\ref{fig:MC1}.

First of all, we notice that for the background-only model (background+DM model) there is a probability of  71\% (55\%) to find $q_\text{bkg} = 0$ ($q_\text{PG} = 0$). This can happen for example when the MC gives a case with no events observed in both LUX and CDMS-Si implying $q_\text{bkg} = q_\text{PG} = 0$.\footnote{Note that, since we do not make any assumption on the background in SuperCDMS, no positive evidence for DM can come from this experiment.} However, we also obtain $q_\text{bkg} \approx 0$ for cases where there are only high-energy events in CDMS-Si, which are firmly excluded by the bound from LUX. Similarly, if there are only events in LUX, we typically find $q_\text{PG} \approx 0$, since neither SuperCDMS nor CDMS-Si can constrain such an excess. 

We furthermore observe from the left panel of Fig.~\ref{fig:MC1} that for $x > 0$ the probability to have $q_\text{bkg} > x$ falls off roughly exponentially with increasing $x$, with some discreet features arising from the binning of the data in LUX.  Based on 5700 data sets, we determine the $p$-value of the background-only hypothesis for the actually observed events ($\bar{q}_\text{bkg} = 5.1$) to be $(1.46 \pm 0.16)\%$. Given the excess of events observed by CDMS-Si it is unsurprising that assuming only known backgrounds the background hypothesis can be rejected with a probability of almost $98.5\%$. The fact that this number is smaller than the value quoted by the CDMS-II collaboration ($99.8\%$) reflects the difficulty to interpret the observed excess in terms of DM when taking the exclusion bounds from LUX and SuperCDMS into account. 

The crucial question therefore is whether the data can be well described by the best-fit DM model. Since we found $\bar{q}_\text{PG} = 12.6 \gg \bar{q}_\text{bkg}$ we expect to find a very small $p$-value for the best-fit point. Indeed, based on 2500 data sets, we find that the probability to have $q_\text{PG} > \bar{q}_\text{PG}$ is only $(0.44 \pm 0.13)\%$. In other words, if the best-fit DM model were the true model of nature, there should be less tension between the individual experiments than observed with a probability of $99.6\%$.

\begin{figure}[b]
\centering
\includegraphics[width=0.46\textwidth]{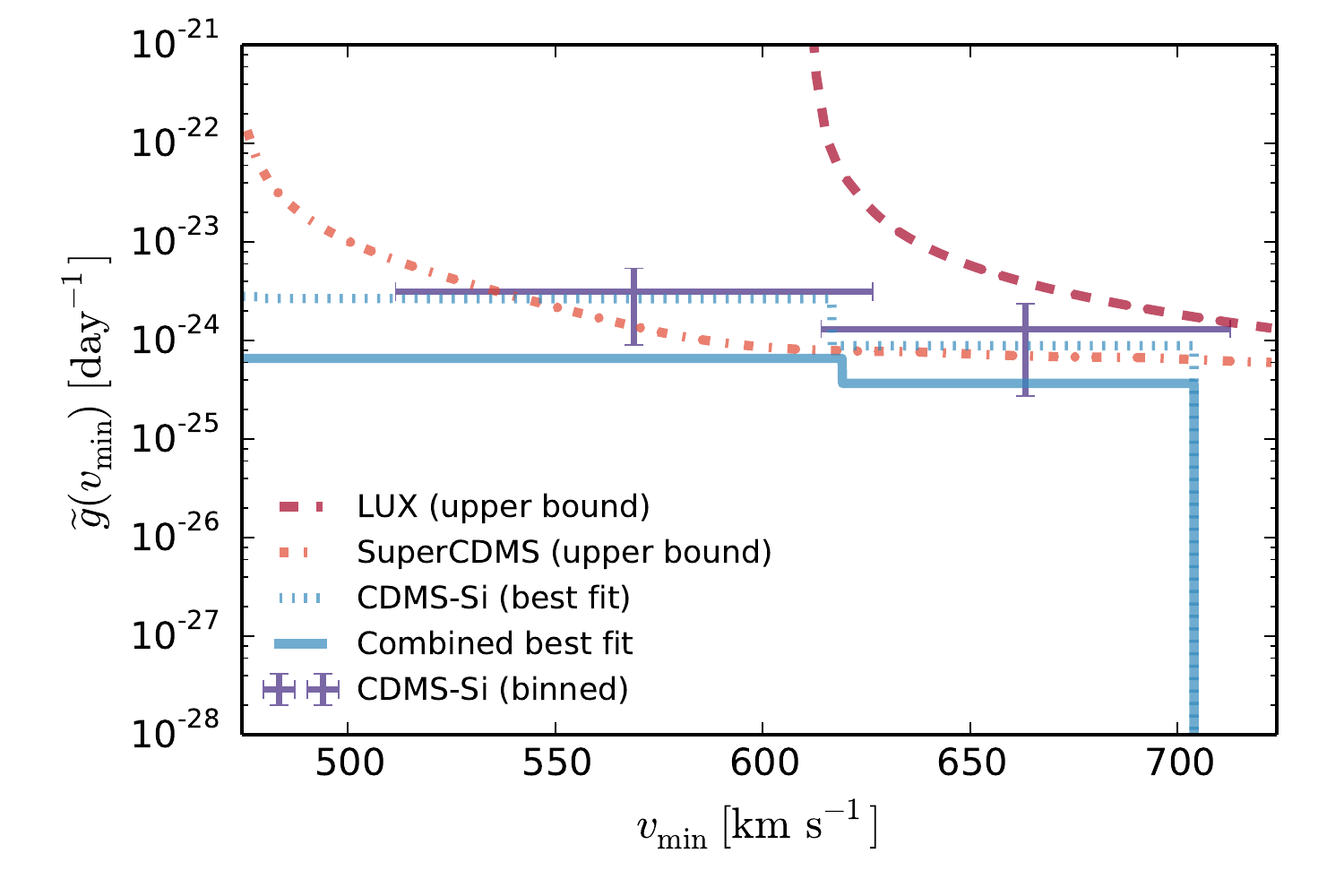}
\quad
\includegraphics[width=0.46\textwidth]{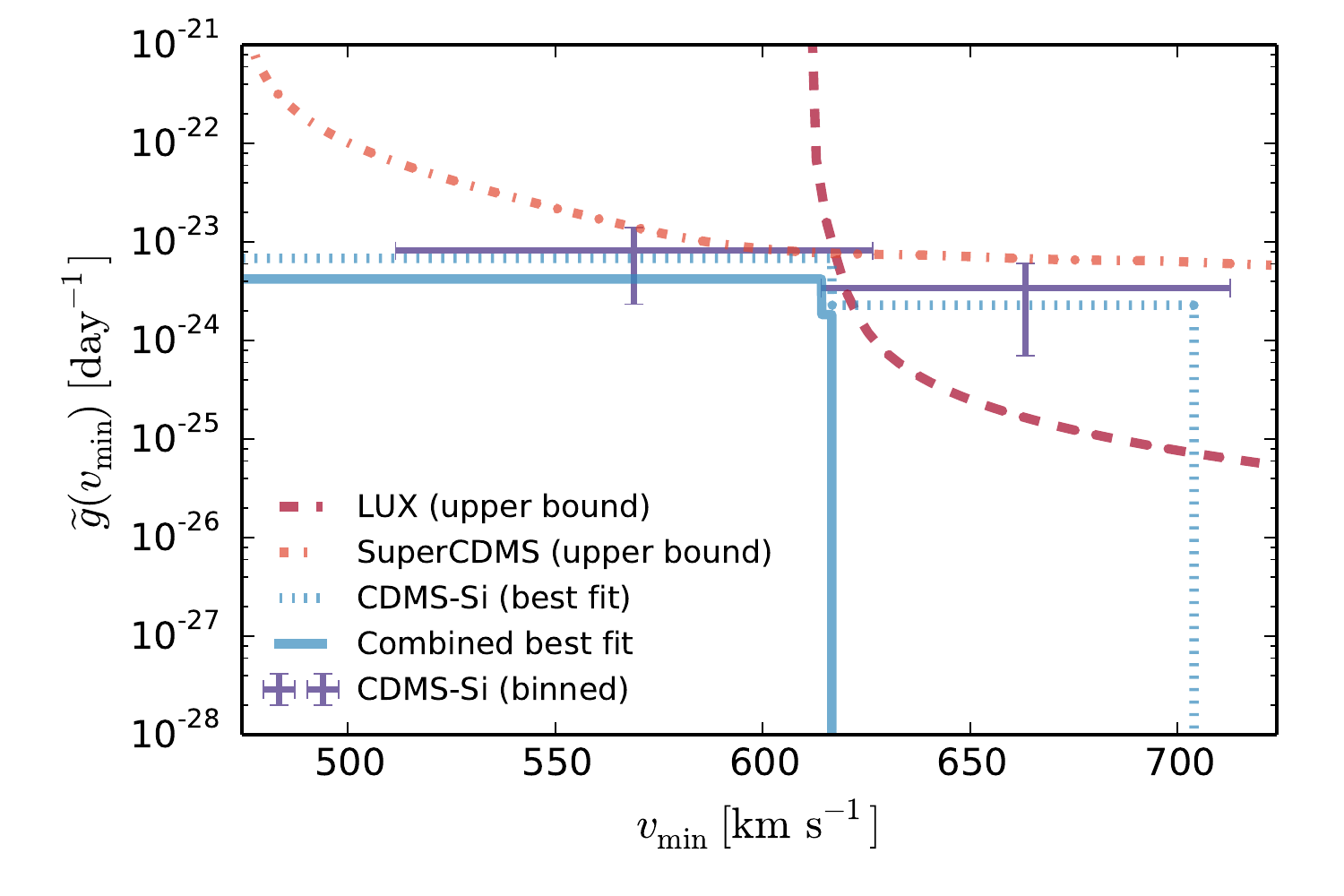}
\caption{Best-fit velocity integral for $f_n / f_p = -0.7$ (left) and $f_n / f_p = -0.8$ (right). In both cases, we have set $m_\chi = 7\:\text{GeV}$.}
\label{fig:nonisoscalar}
\end{figure}

In summary, we find that the known-background-only hypothesis is disfavored with $98.5\%$ probability. Nevertheless, the individual experimental results exclude each other with a probability of $99.6\%$. We would like to emphasize again that in order to derive these results, we have made no assumptions on the DM velocity distribution apart from imposing an observational constraint on the escape velocity. We therefore confirm the conclusion about the incompatibility of LUX, SuperCDMS and CDMS-Si obtained for the Standard Halo Model in a halo-independent way.

\section{Example 2: CDMS-Si with isospin-dependent couplings}\label{Ex2}
\label{sec:ex2}

It is well known that the constraints from SuperCDMS and LUX can be weakened by considering non-isoscalar DM~\cite{Kurylov:2003ra,Giuliani:2005my,Chang:2010yk,Feng:2011vu}. In particular, $f_n / f_p = -0.7$ strongly suppresses the bounds from LUX, while $f_n / f_p \sim -0.8$ strongly suppresses the bound from SuperCDMS~\cite{Gelmini:2014psa}. We illustrate these two possibilities in Fig.~\ref{fig:nonisoscalar} using the mapping into $v_\text{min}$-space introduced in Sec.~\ref{sec:ex1}. Indeed, this figure demonstrates one of the key problems of the conventional presentation of these results: It is impossible to tell by looking at these plots which choice of $f_n / f_p$ actually leads to the better agreement between the three experiments. Similarly, one might wonder whether even better agreement can be obtained for some intermediate value of $f_n / f_p$. By using the global likelihood function we are now in a position to answer these questions.

\begin{figure}[t]
\centering
\includegraphics[width=0.45\textwidth]{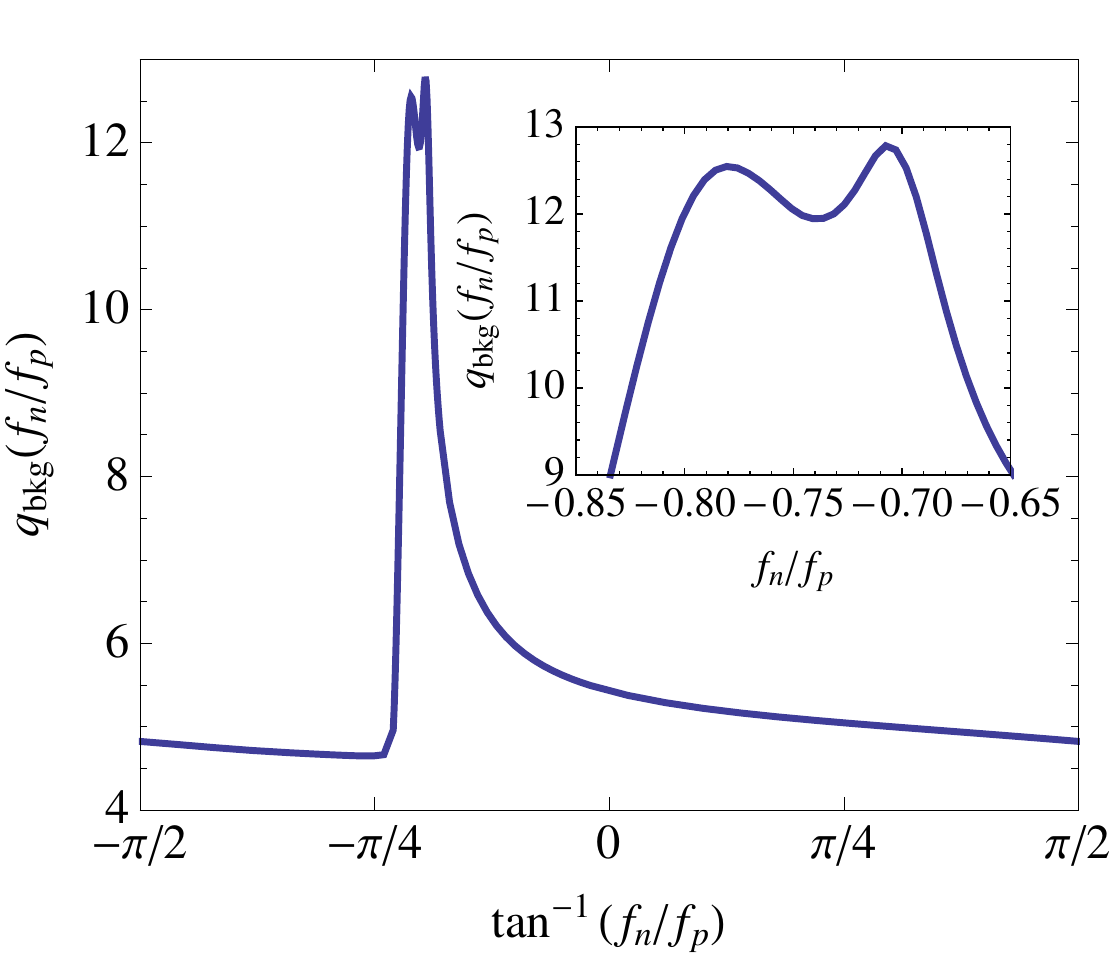}
\quad
\includegraphics[width=0.48\textwidth]{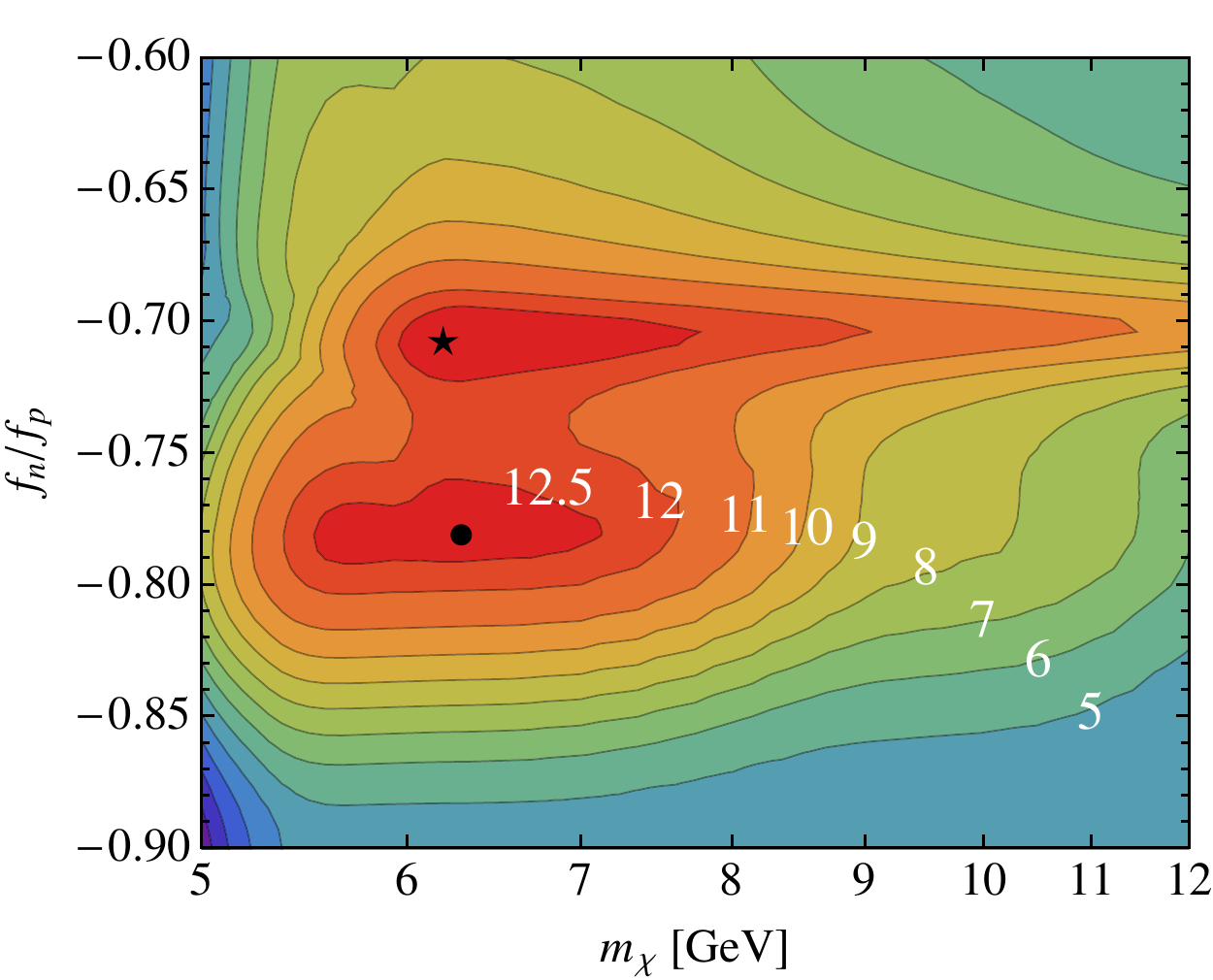}
\caption{The likelihood ratio $q_\text{bkg}$ as a function of $f_n / f_p$ for $m_\chi = 7\:\text{GeV}$ (left) and as a function of both $f_n / f_p$ and $m_\chi$ (right).}
\label{fig:scan}
\end{figure}

\begin{figure}[b]
\centering
\includegraphics[width=0.46\textwidth]{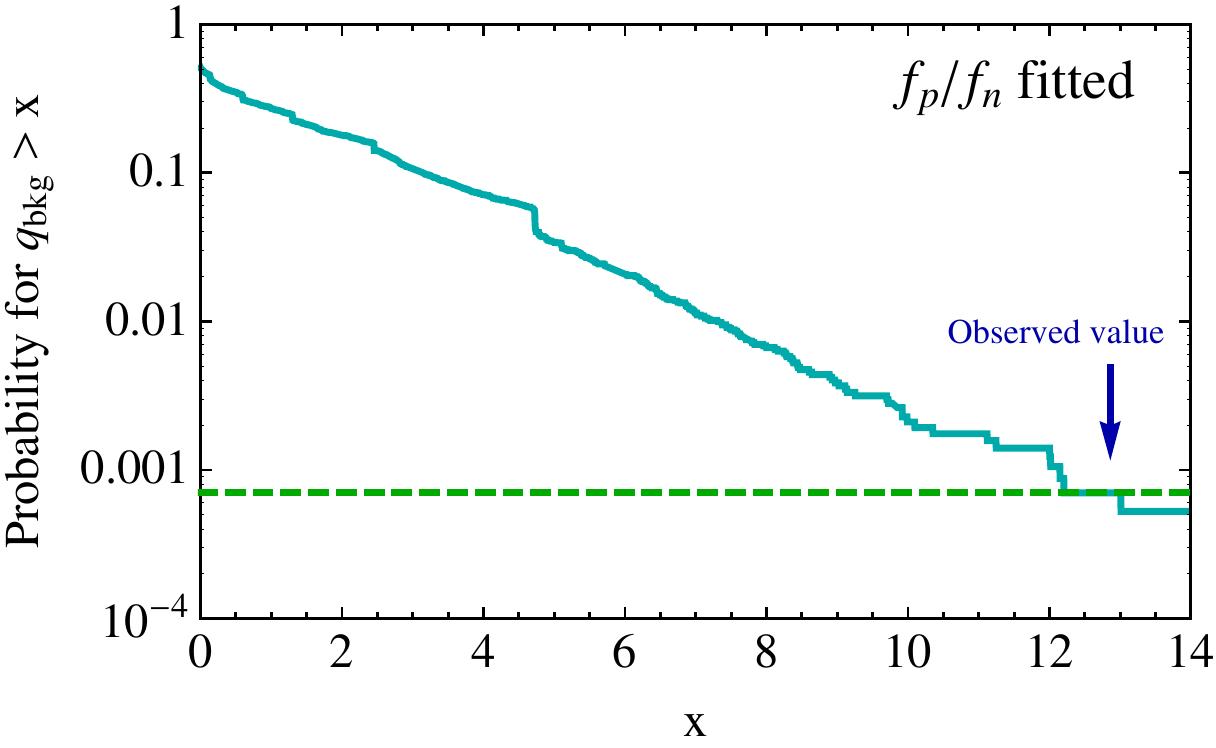}
\caption{$p$-value for $q_\text{bkg}$ based on the background-only hypothesis when fitting both $m_\chi$ and $f_n / f_p$.}
\label{fig:MCBG}
\end{figure}

The left panel of Fig.~\ref{fig:scan} shows the value of $q_\text{bkg} $ as a function of $f_n / f_p$ for $m_\chi = 7$ GeV. This figure allows to directly compare the two cases shown in Fig.~\ref{fig:nonisoscalar}. We observe that there is a slight preference for the suppression of LUX ($f_n / f_p = -0.7$) compared to the suppression of SuperCDMS($f_n / f_p = -0.8$). In the right panel we scan over both $f_n / f_p$ and $m_\chi$ simultaneously. As in Sec.~\ref{sec:ex1}, we have introduced an extra term in the likelihood function to disfavor very small masses and unacceptably large escape velocities. We find the global minimum at $m_\chi = 6.2\;\text{GeV}$ and $f_n / f_p = -0.71$. A second (almost degenerate) minimum is found at $m_\chi = 6.3\:\text{GeV}$ and $f_n / f_p = -0.79$. For these two best-fit points we find $q_\text{bkg} = 12.9$, $q_\text{PG} = 4.8$ and $q_\text{bkg} = 12.6$, $q_\text{PG} = 5.1$ respectively.

As above, we first of all want to determine the $p$-value for the background-only hypothesis. The cumulative distribution function for $q_\text{bkg}$ based on 5700 data sets is shown in Fig.~\ref{fig:MCBG}. First of all, we observe that the probability to have no preference for a DM signal at all is only $49\%$ (compared to 71\% for the case of $f_n = f_p$). The reason is that, by varying $f_n / f_p$, it is now possible to find a consistent DM interpretation for a larger number of upward fluctuations. The observed value of $q_\text{bkg}$ in this case is significantly larger than in the case without isospin-dependence. Consequently, we find a $p$-value for the background-only hypothesis of $p(\text{background}) = (0.070 \pm 0.035) \%$. In other words, the background-only hypothesis is disfavored with a probability of about $99.93\%$.\footnote{This value is insignificantly larger than the one quoted by the CDMS-II collaboration for CDMS-Si alone, since there is a very slight preference for a DM signal also in the data from LUX, contributing to $\bar{q}_\text{bkg}$.}

\begin{figure}[t]
\centering
\includegraphics[width=0.46\textwidth]{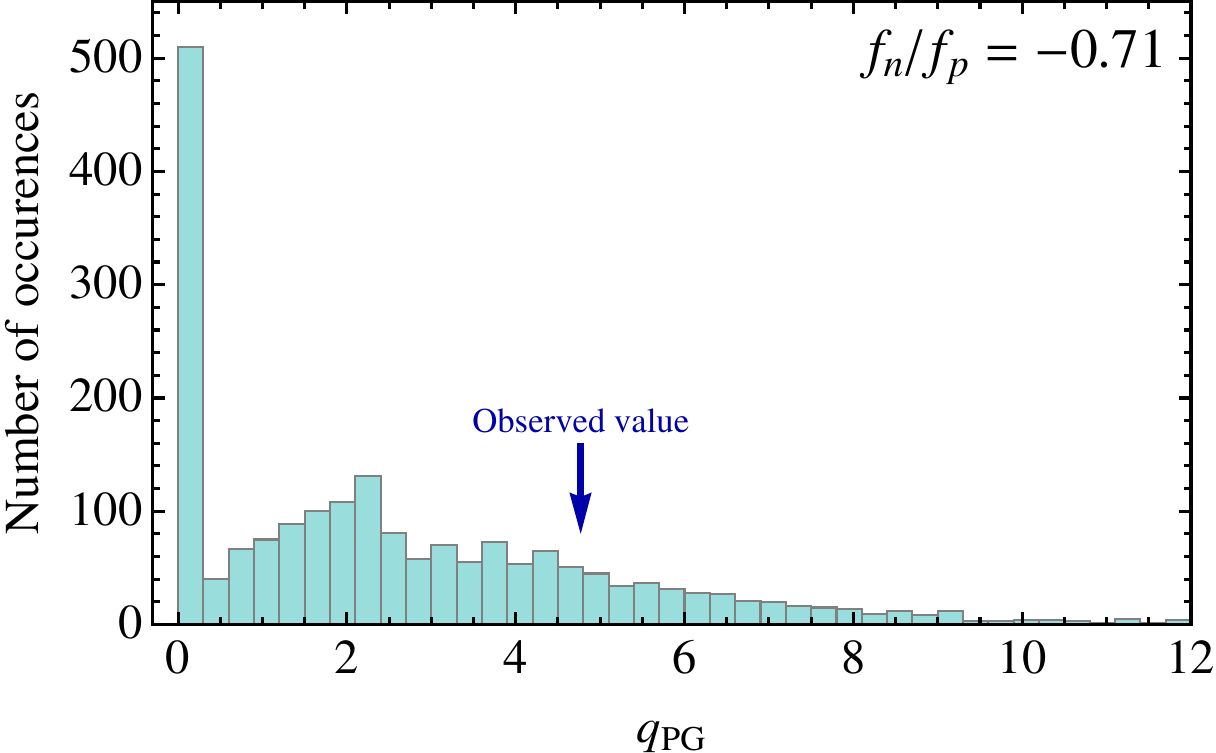}
\hfill
\includegraphics[width=0.46\textwidth]{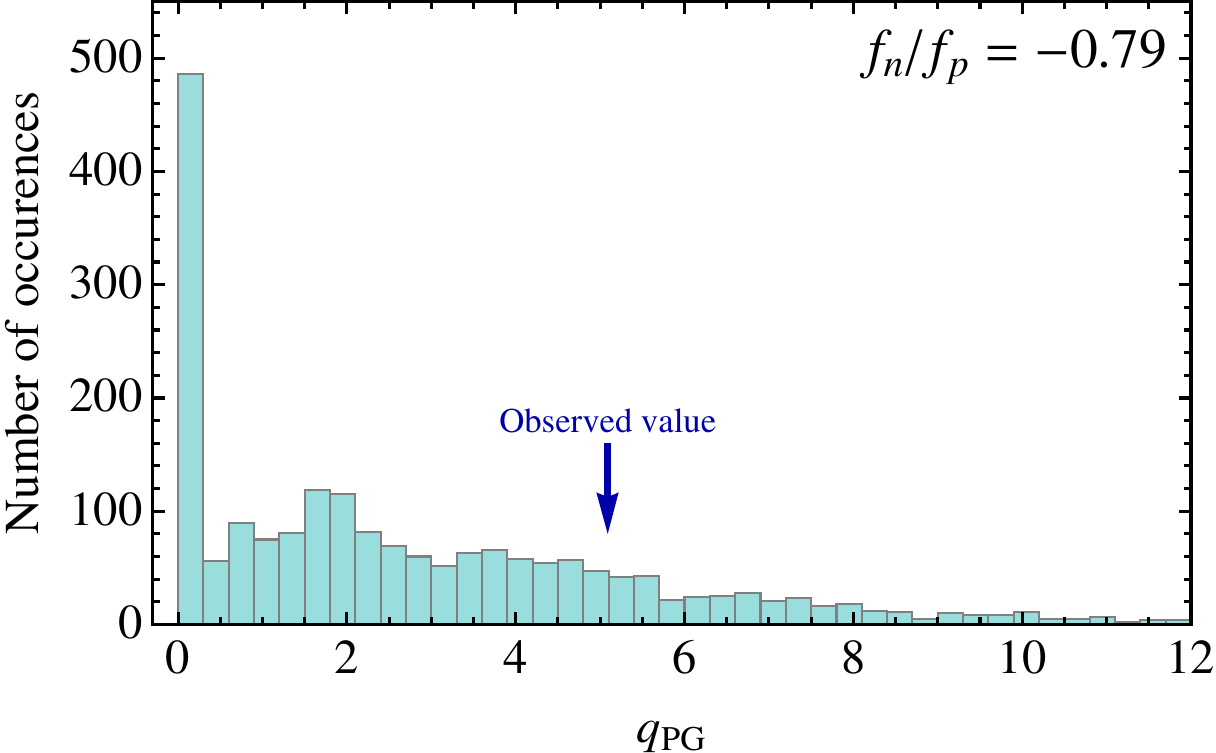}
\caption{The distribution of the test statistic $q_\text{PG}$ based on a sample of 2000 MC simulations of the experiments under consideration. For the left plot we have assumed that the true model of nature corresponds to the global maximum of the likelihood function (suppression of LUX), whereas the right plot corresponds to the second local minimum (suppression of SuperCDMS).}
\label{fig:MC2}
\end{figure}

Nevertheless, as we have seen above it is not sufficient to simply reject the assumption of known backgrounds only. We also have to demonstrate that the proposed DM model provides a good fit to the data in the sense that the tension between the individual experiments is no larger than what is expected from random fluctuations of the data. We have therefore run MC simulations for both minima found in Fig.~\ref{fig:scan} and determined the value of $q_\text{PG}$ for each data set. The results based on 2000 MC samples are shown in Fig.~\ref{fig:MC2}.

We observe that in contrast to the case of $f_n = f_p$ the observed values of $q_\text{PG}$ now lie well within the central regions of the distributions. Indeed, we find that for the global minimum, a value of $q_\text{PG}$ as large as (or larger than) the observed value $\bar{q}_\text{PG} = 4.8$ will occur with a probability of $(18.7 \pm 1.0)\%$. For the second local minimum, the $p$-value is still $(18.5 \pm 1.0)\%$. In other words, the ``tension'' between the different experiments is no larger than what would be expected from random fluctuations of a consistent DM model in $20\%$ of the cases.

\section{Discussion}
\label{sec:discussion}

In this work, we have presented a new quantitative method for analyzing an emerging DM signal from direct detection experiments in a way that is independent of uncertainties in the halo velocity distribution.  This is the first such method in the literature which makes no assumptions about the form of the velocity distribution, and yet still allows for precise conclusions to be drawn  about the significance  of the signal in the data, as well as the compatibility of the results of multiple experiments.  Our method makes use of two test statistics, $q_\text{bkg}$, which characterizes the goodness of fit relative to the background only hypothesis, and $q_\text{PG}$, which evaluates the compatibility of the full set of data. 

We have concentrated on analyzing the best-fit point in DM parameter space (as defined by  $q_\text{PG}$ or equivalently by $q_\text{bkg}$). Where the evidence for a DM signal is found to be significant, one might also be interested in computing confidence intervals in that parameter space. This is feasible with our method as well, though it would likely be time consuming numerically.  The reason is that  for experiments seeing few events we cannot determine the distribution of $q_\text{PG}$ independent of the DM parameters, but rather our method currently requires MC simulations to determine the $p$-values for $q_\text{BKG}$ and $q_\text{PG}$ for each set of model parameters.

In practice, we do not expect the probability distribution for the PGS to exhibit a strong dependence on the DM parameters. For example, for the two different sets of parameters considered in Fig.~\ref{fig:MC2} we obtain almost identical distributions. Nevertheless, it is conceivable that one might find a set of DM parameters that gives a comparably high goodness-of-fit even though it does not minimise the likelihood function. To study these possibilities in detail, it would be highly desirable to have~--- at least approximately~--- an analytical understanding of the distribution functions.

In this context we note that for all the cases considered in this paper, the distribution of $q_\text{bkg}$ has a faster falling tail than the distribution of $q_\text{PG}$. In other words, for comparable values of $q_\text{bkg}$ and $q_\text{PG}$, the latter typically has the larger $p$-value. This observation is related to the fact that we study $q_\text{bkg}$ under the assumption of only background events and $q_\text{PG}$ using the best-fit DM model.\footnote{We have checked that using the background-only model we obtain a distribution for $q_\text{PG}$ that is very similar to the one for $q_\text{bkg}$.} We can therefore conclude that a certain amount of information can already be extracted without the need for MC simulations: If for a given best-fit DM model, we find $q_\text{PG} < q_\text{bkg}$, we can conclude that the tension between the individual experiments is smaller than the tension with the background-only hypothesis. In other words, in such a case the probability is higher for the observed data to have arisen from fluctuations of the DM model than from fluctuations of the background. If, on the other hand, $q_\text{PG}$ is significantly larger than $q_\text{bkg}$, we can conclude that the proposed DM interpretation is not a good explanation for the observed data and that indeed background fluctuations could better account for the experimental results. Only if $q_\text{bkg} \sim q_\text{PG}$, or if a more quantitative statement is required, will MC simulations become necessary.

Finally, we would like to point out an interesting result concerning the overall DM-nucleon scattering cross section. 
For the entire discussion so far, we have absorbed the DM-proton scattering cross section into the definition of the rescaled velocity integral
$\tilde{g}(v_\text{min}) = \frac{\rho \, \sigma_p}{m_\chi}  \int_{v>v_\text{min}}   \hspace{-1mm} f(\mathbf{v}) / v \, \mathrm{d}^3\mathbf{v}$, which is the quantity we have optimized during our analysis.  This has enabled us to vary the overall normalization of $\tilde{g}(v_\text{min})$ arbitrarily, since  $\sigma_p$ is a priori unknown.  However, producing appropriately large values for $\tilde{g}(v_\text{min})$ may require correspondingly large values of $\sigma_p$, which might be possible to test via other types of experiments, such as collider searches~\cite{Chatrchyan:2012me, ATLAS:2012ky}.
Indeed, we have 
\begin{equation}
\int_0^\infty \tilde{g}(v) \mathrm{d}v = 
- \int_0^\infty v \, \tilde{g}'(v) \, \mathrm{d}v = \frac{\rho \sigma_p}{m_\chi}  \int_0^\infty f(v) \, \mathrm{d}v =  \frac{\rho \sigma_p}{m_\chi} \;,
\end{equation}
where $f(v) = \int f(\mathbf{v}) \, v^2 \, \mathrm{d}\Omega_v$. Now, direct detection experiments only allow us to determine $g(v_\text{min})$ in a certain range of velocities $\left[v_\text{low}, v_\text{high}\right]$. In other words, we cannot determine the fraction of the local DM density that has a velocity below $v_\text{low}$ or above $v_\text{high}$. Nevertheless, we can use the monotonicity and positivity of the velocity integral to yield a lower bound on $\sigma_p$ as desired:
\begin{equation}
\int_0^\infty \tilde{g}(v) \geq v_\text{low} \tilde{g}(v_\text{low}) + \int_{v_\text{low}}^{v_\text{high}} \tilde{g}(v_\text{min})
\equiv \tilde{G} \; ,
\end{equation}
where the best fit value of $\tilde{G}$ may be calculated in our method, and we then have
\begin{equation}
\sigma_p \geq \frac{\tilde{G} \, m_\chi}{\rho} \; .
\end{equation}
For the best-fit velocity integral obtained in section \ref{Ex2} for the CDMS-Si, SuperCDMS and LUX data, and assuming $\rho \lesssim 0.3 \: \text{GeV\,cm}^{-3}$ we find approximately $\sigma_p \gtrsim 1.7 \times 10^{-41} \: \text{cm}^2$. It is important to note that this is not a $90\%$ confidence level upper bound on the scattering cross section, but simply the statement that the best-fit velocity integral is mathematically inconsistent with smaller values of $\sigma_p$. Consequently, if complementary searches for DM can exclude scattering cross sections of this magnitude, they would also exclude the best-fit DM interpretation of the current data from direct detection experiments in a halo-independent way.

\section*{Acknowledgements}

We thank Julien Billard, Nassim Bozorgnia, Glen Cowan, Bradley Kavanagh, Christopher McCabe for useful discussions and Thomas Schwetz for valuable comments on the manuscript. BF is supported by STFC UK and FK is supported by a Leathersellers' Company Scholarship at St Catherine's College, Oxford.

\appendix

\section{Constructing the global likelihood function}
\label{ap:experiments}

In this appendix we provide the experimental details needed to construct the likelihood functions that we use for our global analysis.

\subsection*{CDMS-Si}

The CDMS-II collaboration has observed 3 events in the DM search region of their silicon detectors (CDMS-Si for short), compared to a background expectation of only $0.62$ events~\cite{Agnese:2013rvf}. For our analysis we assume an energy resolution of $\Delta E_\text{R} = 0.3\:\text{keV}$~\cite{Akerib:2010pv} and use the detector acceptance from Ref.~\cite{Agnese:2013rvf}. For the background estimate we take the normalized background distributions from Ref.~\cite{McCarthy} and rescale the individual contributions in such a way that 0.41, 0.13 and 0.08 events are expected from surface events, neutrons and $^{206}$Pb, respectively, as in~\cite{Agnese:2013rvf}. The likelihood for a given DM model is then calculated using Eq.~(\ref{eq:unbinnedL}).

\subsection*{SuperCDMS}

We use the results from Ref.~\cite{Agnese:2014aze} based on an exposure of $577\:\text{kg\,days}$. We divide the search region $\left[1.6\:\text{keV},\,10\:\text{keV}\right]$ into 8 evenly spaced bins and calculate the expected number of events in each bin using the detection efficiency from Ref.~\cite{Agnese:2014aze} and assuming a detector resolution of $\Delta E_\text{R} = 0.3\:\text{keV}$ as above. We then calculate the likelihood following Eq.~(\ref{eq:binnedL}). 

Following the analysis by the SuperCDMS collaboration, we do not make any assumption on the level of background. In any bin where the number of events predicted from DM is smaller than the number of observed events, we therefore take the background to account for the difference, such that this bin gives no contribution to $\log \mathcal{L}$. If the number of events predicted from DM is larger than the number of observed events, we take the background contribution to be zero. Effectively this procedure means that only bins where the predicted DM signal exceeds the number of observed events give a finite contribution to $\log \mathcal{L}$. This very conservative approach reproduces the published bound (see Fig.~\ref{fig:standard}), but implies that we can only ever obtain an upper bound, but not positive evidence for DM from SuperCDMS.

\subsection*{LUX}
\label{ap:LUX}

We analyse the results from the first run of LUX, based on an exposure of $10065.4\:\text{kg\,days}$~\cite{Akerib:2013tjd}. In principle, the optimum way to calculate the likelihood for LUX would be to construct an unbinned likelihood function based on the two-dimensional distributions of signal and background in the $S1$-$\log_{10}(S2/S1)$ parameter plane. Unfortunately, not all the necessary information on these distributions is publicly available.
We will therefore define a more restrictive signal region by considering only events that fall below the mean of the nuclear recoil band. Nevertheless, to retain at least some information on the energy of the observed events, we divide the signal region into 8 bins that are evenly spaced in reconstructed recoil energy. In order to calculate the likelihood for LUX we therefore need two key ingredients: First of all, we need to know the acceptance for each bin as a function of the true nuclear recoil energy. In other words, we need to determine the probability that a nuclear recoil with a given recoil energy is reconstructed with a recoil energy within a given bin and has a value of $\log_{10}(S2/S1)$ signal below the mean of the nuclear recoil band. And second, we need to estimate the background expectation in each bin. We will now explain how we extract these two quantities from the publicly available information.

To determine the detector acceptance, we need to estimate the distribution of the $S1$ and $S2$ signals as a function of the true nuclear recoil signal. For the distribution of the $S1$ signal, we take the relative scintillation yield $\mathcal{L}_\text{eff}$ and the absolute yield from~\cite{LUXtalk}. The average detection efficiency for a scintillation photon is 0.14~\cite{Akerib:2013tjd}. For the fluctuations we make the standard assumptions~\cite{Aprile:2011hx} that for an expected $S1$ signal of $\langle S1 \rangle = \nu$ the $S1$ distribution is given by
\begin{equation}
p(S1; E_R) = \epsilon(S1) \, \sum_{n=1}^{\infty} \text{Gauss}(S1; n, \sigma(n)) \, \text{Pois}(n; \nu)
\end{equation}
where $\text{Pois}(n; \nu)$ is a Poisson distribution with expectation value $\nu$ and $\text{Gauss}(S1; n, \sigma(n))$ is a normal distribution with mean $n$ and standard deviation $\sigma(n)$. Because of the similarity of LUX and XENON100, we assume $\sigma(n) = 0.5 \sqrt{n}$ as in~\cite{Aprile:2011hx}. Finally $\epsilon(S1)$ gives the detector acceptance for an $S1$ signal of given magnitude. This quantity has been extracted by comparing calibration data to numerical simulations~\cite{Akerib:2013tjd}.

For the distribution of the $S2$ signal, we take the ionization yield for an electric field of 181 kV/m from~\cite{NEST}. According to~\cite{Akerib:2013tjd} the average probability for an electron to be extracted into the gas phase is 0.65. We include an additional factor of 0.95 to account for the probability that an electron will recombine before it reaches the liquid-gas interface. Once the electron reaches the gas phase, it will create on average a raw S2 signal of 25.8 phe~\cite{LUXtalk}, out of which on average 11.1 are on the bottom PMT array, contributing to the ``$S2_b$" signal.\footnote{The latter number can be reconstructed from observing that the cut $S2 > 200$ phe corresponds to $S2_b > 86$ phe.}  It is the $S2_b$ signal that is used to define the nuclear recoil band by the LUX collaboration, and its total value is then given by the number of electrons created at the interaction point $n_e$ times the amplification factor $f_a \approx 6.9 = 0.65 \times 0.95 \times 11.1$. Taking into account both fluctuations in $n_e$ and fluctuations in the amplification, the $S2_b$ distribution is given by
\begin{equation}
p(S2_b; E_R) = \epsilon(S2_b) \sum_{n=1}^{\infty} \text{Gauss}(S2_b; f_a \, n, \sigma(f_a \, n)) \, \text{Pois}(n; n_e)
\end{equation}
where $\sigma$ is defined as before and $\epsilon(S2_b)$ is the acceptance of the $S2_b$ signal. According to Fig.~6 in~\cite{Akerib:2013tjd} this latter quantity should be very close to unity for the signals we are interested in. Since we have $n_e \gg 1$ for recoil energies above the threshold, we can approximate the Poisson distribution by a Gaussian and, after convolving the two Gaussians, we obtain approximately
\begin{equation}
p(S2_b; E_R) = \text{Gauss}(S2_b; f_a \, n_e, \tilde{\sigma}(n_e))
\end{equation}
where $\tilde{\sigma}(n_e) = \sqrt{0.25 \, f_a \, n_e + f_a^2 \, n_e} \approx f_a \sqrt{n_e}$.

We neglect anti-correlation between the $S1$ and $S2_b$ signals and therefore write the combined distribution as
\begin{equation}
p(S1, S2_b; E_R) = p(S1; E_R)p(S2_b; E_R) \; .
\end{equation}
Convoluting this distribution with a given recoil spectrum then yields the expected $S1$-$S2_b$ distribution. A simple variable transformation to $lS = \log_{10} S2_b/S1$ then gives
\begin{equation}
p(S1, lS; E_R) = p(S1, S2_b; E_R) \, \log 10 \, 10^{lS} \, S1 \; .
\end{equation}

\begin{figure}[t]
\centering
\includegraphics[width=0.48\textwidth]{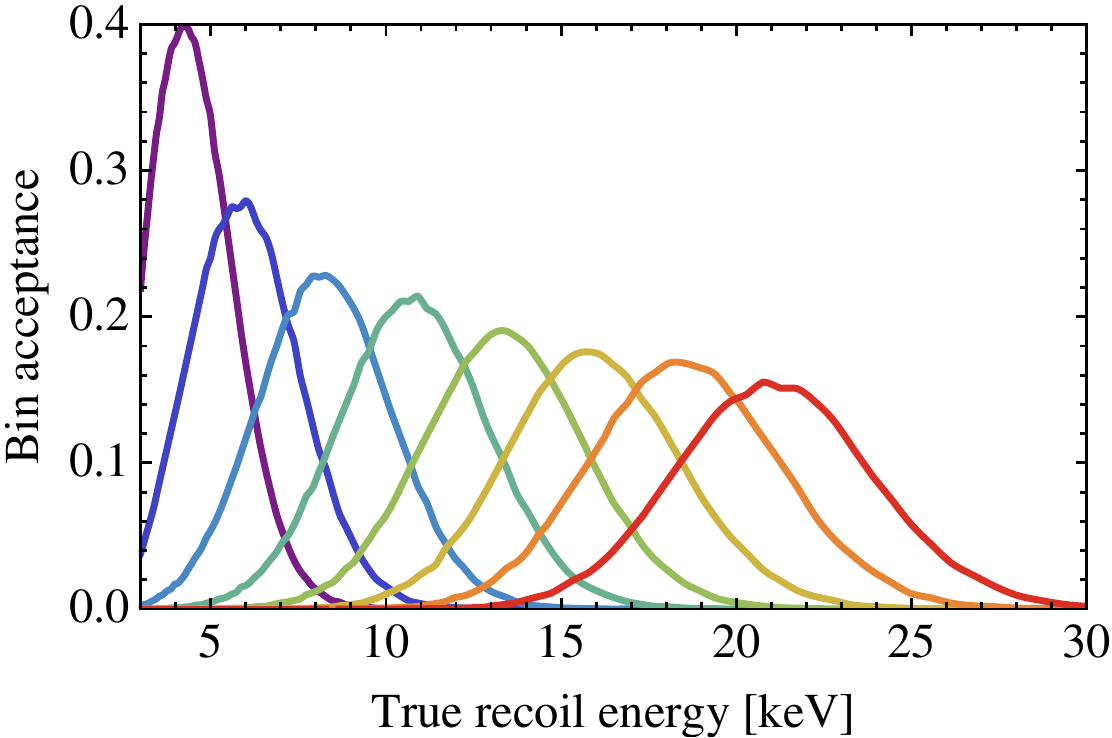}
\caption{The acceptance probability as a function of the true nuclear recoil energy for the 8 different bins under consideration.}
\label{fig:LUXimplementation}
\end{figure}

Having constructed the distributions as above, we can very easily run a MC simulation of DM recoils, in order to generate mock calibration data and validate our implementation. We find that our implementation is sufficient to reproduce the nuclear recoil band from~\cite{Akerib:2013tjd} to good approximation. Similarly, we reproduce the various cut acceptances given in~\cite{Akerib:2013tjd}. Using these data sets we can now extract the acceptance for each bin. The resulting acceptance functions are shown in Fig.~\ref{fig:LUXimplementation}.

Before we can calculate the likelihood for a given DM model, we need an estimate of the background distribution. We assume that the background is dominated by leakage from the electron recoil band and neglect the contribution from neutrons. The $S1$ distribution of electron recoils is given in~\cite{Akerib:2013tjd}. To calculate the number of electron recoils that leak into each bin (i.e. that have a value of $\log_{10} S2_b/S1$ below the mean of the nuclear recoil band), we furthermore need to know the leakage fraction as a function of energy. This quantity was also estimated in~\cite{Akerib:2013tjd}. Combining all available information, we find that the distribution of electron recoils below the nuclear recoil band can be well fitted by $d_\text{bkg}(E_R) = (0.002 \: \text{keV}^{-1}) \cdot (1 + (E_R / \text{keV}))$. This approximation predicts 0.03, 0.04, 0.06, 0.07, 0.09, 0.10, 0.12 and 0.13 background events in the 8 bins under consideration, in agreement with the published prediction of a total background of 0.64 events.

\section{Details on the Monte Carlo simulations}

The MC simulations used to determine the distribution functions for the test statistics $q_\text{bkg}$ and $q_\text{PG}$ are based on a three-step process: The calculation of the expected event rates for a given model, the event generation and the determination of the best-fit DM parameters for each set of events. In this appendix, we will provide some additional information on each of these steps.

The calculation of the expected event rates for a given model proceeds in complete analogy to the calculation used to determine the likelihood for a given model. The only subtlety is that in order to run MC simulations of SuperCDMS, we actually need to make an assumption on the background expectation. To be consistent with the approach outlined in Appendix~\ref{ap:experiments}, we determine the background from the actual observations. In every bin where the model prediction exceeds the number of observed events, we set the background equal to zero, in all other bins we set the background equal to the difference between observation and prediction. This approach means that the MC simulation for SuperCDMS actually depends not only on the assumed model, but also on the experimental data. Given further information on the background expectation in SuperCDMS, it would be straight-forward to implement a different approach.

For SuperCDMS and LUX, where we consider only binned numbers of events, it is straight-forward to generate large samples of events using Poisson distributions centred at the expected number of events. For CDMS-Si, on the other hand, we actually need to sample the full distribution function, which (for a best-fit halo with several steps) can have several complicated features. For this purpose, we calculate the expected energy spectrum by convolving the DM recoil spectrum with the detector acceptance and resolution and then numerically integrate this spectrum to determine the cumulative distribution function (CDF). We can then apply inverse transform sampling, i.e.\ we use the inverse of the CDF and a uniformly distributed sample of random numbers to sample the energy spectrum in CDMS-Si.

Finally, in order to determine $q_\text{bkg}$ and $q_\text{PG}$ for each set of events, we proceed in exactly the same way as for the original data set. In particular this means that we no longer make use of the background estimate in SuperCDMS used to generate the MC sample, and instead determine the best-fit background estimate from the new set of data. Moreover, given the background distributions in LUX and CDMS-Si, we inevitably obtain samples with a number of events at relatively high recoil energies ($E_R>20\:\text{keV}$). This fact, together with our constraint on the escape velocity, means that we can no longer focus just on the low-mass region but need to search for additional minima with large values of $m_\chi$.

Given that, for each MC sample we need to scan over the DM parameters and find the best-fit DM velocity integral at every point, it is clear why it is absolutely crucial to have a very efficient code. Even though we can optimize the halo parameters in a few seconds, a full parameter scan can still take up to 30 minutes. In practice, such an extensive scan is not necessary for each set of events.\footnote{For example, the minimization becomes trivial if in a given sample there are no events observed by CDMS-Si} In the end, a single CPU can generate and process about 100--200 MC samples per day.

\providecommand{\href}[2]{#2}\begingroup\raggedright\endgroup

\end{document}